\documentclass[aps,preprint]{revtex4}%
\usepackage{amsfonts}
\usepackage{amsmath}
\usepackage{amssymb}
\usepackage{graphicx}%
\providecommand{\U}[1]{\protect\rule{.1in}{.1in}}
\newtheorem{theorem}{Theorem}

\newtheorem{remark}[theorem]{Remark}

\begin{document}
\preprint{ }
\title[ ]{Non-Riemannian geometry, Born-Infeld models and trace free gravitational equations}
\author{Diego Julio Cirilo-Lombardo}
\affiliation{Universidad de Buenos Aires, Consejo Nacional de Investigaciones Cientificas y
Tecnicas (CONICET), National Institute of Plasma Physics(INFIP), Facultad de
Ciencias Exactas y Naturales, Cuidad Universitaria, Buenos Aires 1428, Argentina}
\affiliation{Bogoliubov Laboratory of Theoretical Physics, Joint Institute for Nuclear
Research, 141980 Dubna, Russian Federation}

\begin{abstract}
Non-Riemannian generalization of the standard Born-Infeld (BI) Lagrangian is
introduced and analized from a theory of gravitation with dynamical torsion
field. The field equations derived from the proposed action lead to a trace
free gravitational equation (non-riemannian analog to the trace free\ equation
(TFE) from\cite{tf1}\cite{tf2}\cite{tf}) and the field equations for the
torsion respectively. In this theoretical context, the fundamental constants
arise all from the same geometry through geometrical invariant quantities (as
from the curvature $R$). New results involving generation of primordial
magnetic fields and the link with leptogenesis and baryogenesis are presented
and possible explanations given. The physically admisible matter fields can be
introduced in the model via the torsion vector $h_{\mu}$. Such fields include
some dark matter candidates such as axion, right neutrinos and Majorana and
moreover, physical observables as vorticity can be included in the same way.
From a new wormhole soluton in a cosmological spacetime with torsion we also
show that the primordial cosmic magnetic fields can originate from $h_{\mu}$
with the axion field (that is contained in $h_{\mu})$ the responsible to
control the dynamics and stability of the cosmic magnetic field but not the
magnetogenesis itself. The analisys of Grand Unified Theories (GUT)\ in the
context of this model indicates that the group manifold candidates are based
in $SO(10)$, $SU(5)$ or some exceptional groups as $E(6)$,$E\left(  7\right)
$, etc.

\end{abstract}
\volumeyear{year}
\volumenumber{number}
\issuenumber{number}
\eid{identifier}
\date[Date text]{date}
\received[Received text]{date}

\revised[Revised text]{date}

\accepted[Accepted text]{date}

\published[Published text]{date}

\startpage{1}
\endpage{102}
\maketitle
\tableofcontents

\section{Introduction}

The idea to construct a complete geometrization of the physics is very old.
The drawback of the Einstein GR (General Relativity) equations is the RHS:
$R_{\alpha\beta}-\frac{g_{\alpha\beta}}{2}R=\kappa T_{\alpha\beta}$ with the
symmetric tensor (non-geometrical) $\kappa T_{\alpha\beta}$ \ that introduces
\textit{heuristically} the energy-momentum distribution. Similar drawbacks are
contained by the unimodular gravity. It is well known that the unimodular
gravity is obtained from Einstein-Hilbert action in which the
\textit{unimodular condition: }$\sqrt{-\det g_{\mu\nu}}=1$ is also imposed
from the very beginning \cite{tf1}\cite{tf2}\cite{tf}\cite{kkk}. The resulting
field equations correspond to the \textit{traceless} Einstein equations and
can be shown that they are equivalent to the full Einstein equations with the
cosmological constant term $\Lambda$, where $\Lambda$ enters as an integration
constant and the equivalence between unimodular gravity and general relativity
is given by the arbitrary value of lamda. On the other hand the idea that the
cosmological term arises as an integration constant is one of the motivations
for the study of the unimodular gravity, for recent study, see\cite{tfgen}
and\cite{tfsugra} in the context of supergravity. The fact that the
determinant of the metric is fixed has clearly profound consequences on the
structure of given theory. First of all, it reduces the full group of
diffeomorphisms to invariance under the group of unimodular general coordinate
transformations which are transformations that leave the determinant of the
metric unchanged.

Similar thing happens in the non-Riemannian case, as pointed out
in\cite{di}\cite{di1}\cite{di2}\cite{di3}\cite{di4}, where the corresponding
affine geometrical structure induces naturally the following constraint:
$\frac{K}{g}=constant$. This natural constraint impose a condition (ratio)
between both basic tensors through their determinants: the metric determinant
$g$ and the fundamental one $K$ (in the sense of a nonsymmetric theory that
contains the antisymmetric structures), independently of the precise
functional form of $K$ or $g$. In this work our starting point will be
precisely the last one, where a \textit{metric affine structure} in the
space-time manifold (as described in Section II) will be considered. We will
also show that trace free gravitational equations can be naturally obtained
when the Lagrangian function (geometrical action) is taken as a measure
involving a particular combination of the fundamental tensors of the geometry:%
\[
\sqrt{\left\vert \det f\left(  g_{\mu\nu},f_{\mu\nu,}R_{\mu\nu}\right)
\right\vert }%
\]
with the (0,2) tensors $g_{\mu\nu},f_{\mu\nu,}R_{\mu\nu}:$ the symmetric
metric, the antisymmetric one(that acts as potential of the torsion field) and
the generalized Ricci tensor (proper of the non Riemannian geometry). The
three tensors are related with a Clifford structure of the tangent space (for
details see\cite{die5}) where the explicit choice for $f\left(  g_{\mu\nu
},f_{\mu\nu,}R_{\mu\nu}\right)  $ is given in Section III. This type of
Lagrangians, because are non-Riemannian generalizations of \ the well known
Nambu-Goto and Born-Infeld (BI) ones, can be physically and geometrically
analyzed. Due the pure geometrical structure of the theory, induced energy
momentum tensors and fundamental constants (actually functions)
\textit{emerge} naturally. Consequently, this fact allows the physical
realization of the Mach principle that is briefly treated in SectionVIII after
the (trace free) dynamic equations in Section IV are obtained.

In Section V the\textit{ }trace free gravitational equations and the meaning
of \ the cosmological term as integration constant are discussed from the
physical point of view, meanwhile in Section VI the constancy of G (Newton
constant) is similarly discussed. The important role played by the dual of the
torsion field as geometrical energy-momentum tensor is given in Section VII.
Some physical consequences of the model, as the geometrical origin of the
$\alpha\Omega-$dynamo, is presented in Section IX that it is very important
because establish the link between the mathematical structure of the model of
the first part of the article and the physics of the early universe and the
particle physics of the second half of this work. In Section X the direct
relation between the torsion with axion electrodynamics and Chern-Simons
(CS)\ theory is discussed considering the geometrical structure of the dual
vector of the torsion field. In Section XI an explanation about the
magnetogenesis in FRW scenario, the structure of the GUT\ where the SM\ is
derived and the role of the axion in the dynamics of the cosmic magnetic field
is presented. Finally some concluding remarks are given in Section XII.

\section{Basis of the metrical-affine geometry}

The starting point is a hypercomplex construction of the (metric compatible)
spacetime manifold \cite{die5}\cite{mc}
\begin{equation}
M,\,g_{\mu\nu}\equiv e_{\mu}\cdot e_{\nu} \label{eq1}%
\end{equation}
where for each point $p\in M$ there exists a local affine space $A$. The
connection over $A$, $\widetilde{\Gamma}$, define a generalized affine
connection $\Gamma$ on $M$, specified by $\left(  \nabla,K\right)  $, where
$K$ is an invertible $\left(  1,1\right)  $ tensor over $M.$ We will demand
for the connection to be compatible and rectilinear, that is%
\begin{equation}
\nabla K=KT,\qquad\nabla g=0 \label{eq2}%
\end{equation}
where $T$ is the torsion, and $g$ the space-time metric (used to raise and
lower the indices and determining the geodesics), that is preserved under
parallel transport. This generalized compatibility condition ensures that the
generalized affine connection $\Gamma$ maps autoparallels of $\Gamma$ on $M$
into straight lines over the affine space $A$ (locally). The first equation
above is equal to the condition determining the connection in terms of the
fundamental field in the $UFT$ non-symmetric. Hence, $K$ can be identified
with the fundamental tensor in the non-symmetric fundamental theory. This fact
gives us the possibility to restrict the connection to a
\mbox{(anti-)Hermitian} theory.

The covariant derivative of a vector with respect to the generalized affine
connection is given by%
\begin{align}
\label{eq3}\left.  A^{\mu}\right.  _{;\nu}\equiv\left.  A^{\mu}\right.
,_{\nu}+\Gamma_{\ \ \alpha\nu}^{\mu}A^{\alpha}\\
A_{\mu;\nu}\equiv\left.  A_{\mu}\right.  ,_{\nu}-\Gamma_{\ \ \mu\nu}^{\alpha
}A_{\alpha}%
\end{align}

The generalized compatibility condition (\ref{eq2}) determines the $64$
components of the connection by the $64$ equations
\begin{equation}
K_{\nu;\alpha}^{\mu}=K_{\rho}^{\mu}T_{\ \ \nu\alpha}^{\rho}\quad\text{ where
}\quad T_{\ \ \nu\alpha}^{\rho}\;\equiv2\Gamma_{\ \ \left[  \alpha\nu\right]
}^{\rho} \label{eq4}%
\end{equation}
Notice that by contracting indices $\nu$ and $\alpha$ in the first equation
above, an additional condition over this hypothetical fundamental
(nonsymmetric) tensor $K$ is obtained%
\[
K_{\mu\alpha;}^{\text{ \ \ \ }\alpha}=0
\]
that, geometrically speaking, reads
\[
d^{\ast}K=0.
\]
This is a current-free condition over the tensor $K$. Notice that the metric
is used here to down the indices (metric compatible space-time) and
consequently we can work also with $K_{\alpha\nu}=g_{\alpha\beta}K_{\nu
}^{\beta}$

The metric is uniquely determined by the metricity condition, which puts $40$
restrictions on the derivatives of the metric
\begin{equation}
g_{\mu\nu,\rho}=2\Gamma_{\left(  \mu\nu\right)  \rho} \label{eq5}%
\end{equation}
The space-time curvature tensor, that is defined in the usual way, has two
possible contractions: the Ricci tensor $R_{\mu\lambda\nu}^{\lambda}=R_{\mu
\nu}$, and the second contraction $R_{\lambda\mu\nu}^{\lambda}=2\Gamma
_{\ \ \lambda\left[  \nu,\mu\right]  }^{\lambda}$, which is identically zero
due to the metricity condition (\ref{eq2}).

In order to find a symmetry of the torsion tensor, let us denote the inverse
of $K$ by $\widehat{K}$. Therefore, $\widehat{K}$ is uniquely specified by
condition $\widehat{K}^{\alpha\rho}$ $K_{\alpha\sigma}=K^{\alpha\rho}%
\widehat{K}_{\alpha\sigma}=\delta_{\sigma}^{\rho}$.

As it was pointed out in \cite{di}\cite{di1}\cite{di2}\cite{di3}\cite{di4},
inserting explicitly the torsion tensor as the antisymmetric part of the
connection in (\ref{eq4}), and multiplying by $\tfrac{1}{2}\widehat{K}%
^{\alpha\nu},$ results, after straighforward computations, in%
\begin{equation}
\left(  Ln\sqrt{-K}\right)  ,_{\mu}-\Gamma_{\left(  \mu\nu\right)  }^{\nu}=0
\label{eq6}%
\end{equation}
where $K=\det$ $\left(  K_{\mu\rho}\right)  $. Notice that from expression
(\ref{eq6}) we arrive at the relation between the determinants $K$ and $g$:
\[
\frac{K}{g}=\text{constant}%
\]
(strictly a constant scalar function of the coordinates). Now we can write
\begin{equation}
\Gamma_{\ \alpha\nu,\beta}^{\nu}-\Gamma_{\ \beta\nu,\alpha}^{\nu}%
=\Gamma_{\ \nu\beta,\alpha}^{\nu}-\Gamma_{\ \nu\alpha,\beta}^{\nu}\,,
\label{eq7}%
\end{equation}
as the first term of (\ref{eq6}) is the derivative of a scalar. Then, the
torsion tensor has the symmetry%
\begin{equation}
T_{\ \nu\left[  \beta,\alpha\right]  }^{\nu}=T_{\ \nu\left[  \alpha
,\beta\right]  }^{\nu}=0 \label{eq8}%
\end{equation}
This implies that the trace of the torsion tensor, defined as $T_{\ \nu\alpha
}^{\nu}$, is the gradient of a scalar field%
\begin{equation}
T_{\alpha}=\nabla_{\alpha}\phi\tag{10}%
\end{equation}

In reference \cite{mc}an interesting geometrical analysis is presented of
non-symmetric field structures. There, expressions precisely as (\ref{eq1})
and (\ref{eq2}) ensure that the basic non-symmetric field structures
(\emph{i.e.} $K$) take on a definite geometrical meaning when interpreted in
terms of affine geometry. Notice that the tensor K carries the 2-form
(bivector) that will be associated with the fundamental antisymmetric form in
the next Sections. Such antisymmetric form is introduced from the tangent
space via the generalization of the Ambrose-Singer theorem by exponentiation.

\section{Geometrical Lagrangians: the generalized Born-Infeld action}

Let us start with the geometrical Lagrangian introduced in [\cite{di}%
\cite{di1}\cite{di2}\cite{di3}\cite{di4}]%

\begin{equation}
\mathcal{L}_{g}=\sqrt{\det\left[  \lambda\left(  g_{\alpha\beta}%
+F_{\alpha\beta}\right)  +R_{\alpha\beta}\right]  } \tag{11}%
\end{equation}
it can be rewritten as
\begin{equation}
\mathcal{L}_{g}=\sqrt{\det\left(  G_{\alpha\beta}+\mathcal{F}_{\alpha\beta
}\right)  } \tag{12}%
\end{equation}
with the following redefinitions
\begin{equation}
G_{\alpha\beta}=\lambda g_{\alpha\beta}+R_{\left(  \alpha\beta\right)  }\text{
\ \ \ and \ \ }\mathcal{F}_{\alpha\beta}=\lambda F_{\alpha\beta}+R_{\left[
\alpha\beta\right]  } \tag{13}%
\end{equation}
where a totally antisymmetric torsion tensor $T_{\gamma\beta}^{\alpha}=$
$\varepsilon_{\gamma\beta\delta}^{\alpha}h^{\delta}$ is assumed($h^{\delta}$
its dual vector field). Notice that the antisymmetric tensor $F_{\alpha\beta
},$that takes the role of the electromagnetic field, is proportional to the
dual of the potential for the (totally antisymmetric) torsion field\cite{di}%
\cite{di1}\cite{di2}\cite{di3}\cite{di4}. A brief review on the origin of this
type of Lagrangians in the context of unified theories in reductive geometries
is in Appendix I. Consequently the generalized Ricci tensor splits into a
symmetric and antisymmetric part, namely:
\[
R_{\mu\nu}=\overset{R_{\left(  \mu\nu\right)  }}{\overbrace{\overset{\circ}%
{R}_{\mu\nu}-T_{\mu\rho}^{\ \ \ \alpha}T_{\alpha\nu}^{\ \ \ \rho}}}%
+\overset{R_{\left[  \mu\nu\right]  }}{\overbrace{\overset{\circ}{\nabla
}_{\alpha}T_{\mu\nu}^{\ \ \ \alpha}}}%
\]
where $\overset{\circ}{R}_{\mu\nu}$ is the general relativistic Ricci tensor
constructed with the Christoffel connection. The expansion of the determinant
leads to the Born-Infeld generalization in the usual form :%
\begin{align}
\mathcal{L}_{g}  &  =\sqrt{\left\vert G\right\vert }\sqrt{1+\frac{1}%
{2}\mathcal{F}_{\mu\nu}\mathcal{F}^{\mu\nu}-\frac{1}{16}\left(  \mathcal{F}%
_{\mu\nu}\widetilde{\mathcal{F}}^{\mu\nu}\right)  ^{2}}\tag{14}\\
&  =\Lambda^{2}\sqrt{\left\vert g\right\vert }\sqrt{1+\frac{1}{2}\Lambda
_{1}^{2}F_{\mu\nu}F^{\mu\nu}-\frac{1}{16b^{4}}\left(  \Lambda_{2}^{2}F_{\mu
\nu}\widetilde{F}^{\mu\nu}\right)  ^{2}} \tag{15}%
\end{align}
where
\begin{align}
\Lambda &  =\lambda+\frac{g_{\alpha\beta}R^{\left(  \alpha\beta\right)  }}%
{4}\tag{16}\\
\Lambda_{1}^{2}  &  =\lambda^{2}\left(  1+\frac{2}{\lambda}\overset{}%
{\frac{F_{\mu\nu}R^{\left[  \mu\nu\right]  }}{F_{\mu\nu}F^{\mu\nu}}}+\frac
{1}{\lambda^{2}}\frac{R_{\left[  \mu\nu\right]  }R^{\left[  \mu\nu\right]  }%
}{F_{\mu\nu}F^{\mu\nu}}\right) \tag{17}\\
\Lambda_{2}^{2}  &  =\lambda^{2}\left(  1+\frac{2}{\lambda}\overset{}%
{\frac{F_{\mu\nu}\widetilde{R}^{\left[  \mu\nu\right]  }}{F_{\mu\nu}%
\widetilde{F}^{\mu\nu}}}+\frac{1}{\lambda^{2}}\frac{R_{\left[  \mu\nu\right]
}\widetilde{R}^{\left[  \mu\nu\right]  }}{F_{\mu\nu}\widetilde{F}^{\mu\nu}%
}\right)  \tag{18}%
\end{align}
Although the action is exact and have the correct limit, the analysis can be
simplest and substantially improved using the following action%
\begin{align}
\mathcal{L}_{gs}  &  =\sqrt{\det\left[  \lambda g_{\alpha\beta}\left(
1+\frac{R_{s}}{4\lambda}\right)  +\lambda F_{\alpha\beta}\left(  1+\frac
{R_{A}}{\lambda}\right)  \right]  }\tag{19}\\
R_{s}  &  \equiv g^{\alpha\beta}R_{\left(  \alpha\beta\right)  };\text{
\ \ \ \ \ \ \ \ \ }R_{A}\equiv f^{\alpha\beta}R_{\left[  \alpha\beta\right]
}\text{ \ \ \ \ } \tag{20}%
\end{align}
$\left(  with\text{ }f^{\alpha\beta}\equiv\frac{\partial\ln\left(  \det
F_{\mu\nu}\right)  }{\partial F_{\alpha\beta}},\det F_{\mu\nu}=2F_{\mu\nu
}\widetilde{F}^{\mu\nu}\right)  $ that contains all necessary information and
is more suitable to manage. \textit{ }If the induced structure from the
tangent space $T_{p}\left(  M\right)  $(via Ambrose-Singer theorem) is
intrinsically related to a (super)manifold structure, we have seen that there
exists a transformation \cite{weyl}\cite{die5}$U_{A}^{B}\left(  P\right)
=\delta_{A}^{B}+\mathcal{R}_{A\mu\nu}^{B}dx^{\mu}\wedge dx^{\nu}%
\rightarrow\delta_{A}^{B}+\omega^{k}\left(  \mathcal{T}_{k}\right)  _{A}^{B}$(
with A,B.... generally a multi-index) having the same form as the blocks
inside of the square root proposed Lagrangian (19): e.g.$\lambda
g_{\alpha\beta}\left(  1+\frac{R_{s}}{4\lambda}\right)  \sim$ where where the
Poisson structure is evident (as the dual of the Lie algebra of the group
manifold) in our case leading the identification between the group structure
of the tangent space with the space-time curvature as $\mathcal{R}_{A\mu\nu
}^{B}dx^{\mu}\wedge dx^{\nu}\equiv\omega^{k}\left(  \mathcal{T}_{k}\right)
_{A}^{B}$ .

\section{Field equations}

The geometry of the space-time Manifold is to be determined by the Noether
symmetries%
\begin{equation}
\frac{\delta L_{G}}{\delta g^{\mu\nu}}=0,\text{ \ \ \ }\frac{\delta L_{G}%
}{\delta f^{\mu\nu}}=0 \tag{21}%
\end{equation}
where the functional (Hamiltonian) derivatives in the sense of Palatini (in
this case with respect to the potentials), are understood. The choice
"measure-like" form for the geometrical Lagrangian $L_{G}$ (reminiscent of a
nonlinear sigma model), as is evident, satisfy the following principles:

\textit{i) the principle of the natural extension of the Lagrangian density as
square root of the fundamental line element containing also }$F_{\mu\nu}.$

\textit{ii) the symmetry principle between }$g_{\mu\nu}$\textit{ and }%
$F_{\mu\nu}$\textit{(e.g. }$g_{\mu\nu}$\textit{ and }$F_{\mu\nu}%
$\textit{should enter into }$L_{G}$\textit{ symmetrically)}

\textit{iii)the principle that the spinor symmetry, namely}%
\begin{equation}
\nabla_{\mu}g_{\lambda\nu}=0,\text{ \ \ \ \ \ \ \ \ \ }\nabla_{\mu}%
\sigma_{\lambda\nu}=0 \tag{22,23}%
\end{equation}
\textit{with}%
\begin{equation}
g_{\lambda\nu}=\gamma_{\lambda}\cdot\gamma_{\nu}\text{,
\ \ \ \ \ \ \ \ \ \ \ \ \ \ \ \ \ \ }\sigma_{\lambda\nu}=\gamma_{\lambda
}\wedge\gamma_{\nu}\sim\ast F_{\lambda\nu} \tag{24,25}%
\end{equation}
\textit{should be derivable from }$L_{G}$\textit{(21)}

The last principle is key because it states that the spinor invariance of the
fundamental space-time structure should be derivable from the dynamic
symmetries given by (21). The fact that the $L_{G}$ satisfies the 3 principles
shows also that it has the simpler form.

Notice that the action density proposed by Einstein in \cite{ein}in his
nonsymmetric field theory satisfies i) and ii) but not iii).

\begin{remark}
Due the totally antisymmetric character of the torsion field it is completely
determined by the fundamental (structural 2-form) antisymmetric tensor, and
consequently the variations must acquire the form given by expression
(21)\textbf{: metric and torsion have each one their respective potentials
that are in coincidence with the fundamental structure of the geometry. }
\end{remark}

\subsection{$\delta_{g}L_{G}$}

The starting point for the metrical variational procedure is in the same way
as in the standard Born-Infeld theory: from the following factorization of the
geometrical Lagrangian :%
\begin{equation}
\mathcal{L}=\sqrt{\left\vert g\right\vert }\sqrt{\det\left(  \alpha
\lambda\right)  }\sqrt{1+\frac{1}{2b^{2}}F_{\mu\nu}F^{\mu\nu}-\frac{1}%
{16b^{4}}\left(  F_{\mu\nu}\widetilde{F}^{\mu\nu}\right)  ^{2}}\equiv
\sqrt{\left\vert g\right\vert }\sqrt{\det\left(  \alpha\lambda\right)
}\mathbb{R} \tag{26}%
\end{equation}
where
\begin{equation}
b=\frac{\alpha}{\beta}=\frac{1+\left(  R_{S}/4\lambda\right)  }{1+\left(
R_{A}/4\lambda\right)  },\text{ \ \ \ \ \ }R_{S}=g^{\alpha\beta}R_{\alpha
\beta},\text{ \ \ \ \ }R_{A}=f^{\alpha\beta}R_{\alpha\beta},\text{ \ \ }
\tag{27,28,29}%
\end{equation}
and $\lambda$ an arbitrary constant we perform the variational metric
procedure with the following result (details see Appendix II)%
\begin{gather}
\delta_{g}\mathcal{L}=0\Rightarrow R_{\left(  \alpha\beta\right)  }%
-\frac{g_{\alpha\beta}}{4}R_{s}=\frac{R_{s}}{2\mathbb{R}^{2}\alpha^{2}}\left[
F_{\alpha\lambda}F_{\beta}^{\ \lambda}-F_{\mu\nu}F^{\mu\nu}\frac{R_{\left(
\alpha\beta\right)  }}{R_{s}}\right]  +\tag{30, 31}\\
+\frac{R_{s}}{4\mathbb{R}^{2}\alpha^{2}b^{2}}\left[  F_{\mu\nu}\widetilde
{F}^{\mu\nu}\left(  \frac{F_{\eta\rho}\widetilde{F}^{\eta\rho}}{8}%
g_{\alpha\beta}-F_{\alpha\lambda}\widetilde{F}_{\beta}^{\ \lambda}\right)
+\frac{F_{\eta\rho}\widetilde{F}^{\eta\rho}}{2}\frac{R_{\left(  \alpha
\beta\right)  }}{R_{s}}\right]  +\nonumber\\
+2\lambda\left[  g_{\alpha\beta}+\frac{1}{\mathbb{R}^{2}\alpha^{2}}\left(
F_{\alpha\lambda}F_{\beta}^{\ \lambda}+\frac{F_{\mu\nu}\widetilde{F}^{\mu\nu}%
}{2b^{2}}\left(  \frac{F_{\eta\rho}\widetilde{F}^{\eta\rho}}{8}g_{\alpha\beta
}-F_{\alpha\lambda}\widetilde{F}_{\beta}^{\ \lambda}\right)  \right)  \right]
,\nonumber
\end{gather}

\begin{remark}
Notice that:

1) The eq. (31) is\emph{ trace-free} type, consequently the trace of the third
term of the above equation ( that is the cosmological one ) is equal to zero.
This happens trivially if $\lambda=0$ or $\ 4\mathbb{R}^{2}\alpha^{2}=-\left(
F_{\alpha\lambda}F^{\ \alpha\lambda}-\frac{\left(  F_{\mu\nu}\widetilde
{F}^{\mu\nu}\right)  ^{2}}{4b^{2}}\right)  $ In terms of the Maxwell
Lagrangian we have $\left(  \mathbb{R}\alpha\right)  ^{2}=\left(
L_{Maxwell}+\frac{\left(  F_{\mu\nu}\widetilde{F}^{\mu\nu}\right)  ^{2}%
}{16b^{2}}\right)  \equiv\mathcal{W}\left(  I_{S},I_{P},b\right)  $ that allow
us to simplify the eq. (31) once more as follows%
\begin{gather*}
R_{\left(  \alpha\beta\right)  }-\frac{g_{\alpha\beta}}{4}R_{s}=\frac{R_{s}%
}{2\mathcal{W}}\left[  F_{\alpha\lambda}F_{\beta}^{\ \lambda}-F_{\mu\nu}%
F^{\mu\nu}\frac{R_{\left(  \alpha\beta\right)  }}{R_{s}}\right]  +\\
+\frac{R_{s}}{4\mathcal{W}b^{2}}\left[  F_{\mu\nu}\widetilde{F}^{\mu\nu
}\left(  \frac{F_{\eta\rho}\widetilde{F}^{\eta\rho}}{8}g_{\alpha\beta
}-F_{\alpha\lambda}\widetilde{F}_{\beta}^{\ \lambda}\right)  +\frac
{F_{\eta\rho}\widetilde{F}^{\eta\rho}}{2}\frac{R_{\left(  \alpha\beta\right)
}}{R_{s}}\right]  +\\
+2\lambda\left[  g_{\alpha\beta}+\frac{1}{\mathcal{W}}\left(  F_{\alpha
\lambda}F_{\beta}^{\ \lambda}+\frac{F_{\mu\nu}\widetilde{F}^{\mu\nu}}{2b^{2}%
}\left(  \frac{F_{\eta\rho}\widetilde{F}^{\eta\rho}}{8}g_{\alpha\beta
}-F_{\alpha\lambda}\widetilde{F}_{\beta}^{\ \lambda}\right)  \right)  \right]
,
\end{gather*}
2) $b$ takes the place of limiting parameter (maximum value) for the
electromagnetic field strength.

3)$b$ is not a constant in general, in sharp contrast with the Born-Infeld or
string theory cases.

4) Because $b$ is the ratio $\frac{\alpha}{\beta}=\frac{1+\left(
R_{S}/4\lambda\right)  }{1+\left(  R_{A}/\lambda4\right)  }$ involving both
curvature scalars from the contractions of the generalized Ricci tensor: it is
preponderant when the symmetrical contraction of $R_{\alpha\beta}$ is greater
than the skew one.

5) The fact pointed out in ii), namely that the curvature scalar plays the
role as some limiting parameter of the field strength, was conjectured by
Mansouri in \cite{man} in the context of gravity theory over group manifold
(generally with symmetry breaking). In such a case, this limit was established
after the explicit integration of the internal group-valuated variables that
is not our case here.

6) In similar form that the Eddington conjecture: $R_{\left(  \alpha
\beta\right)  }\varpropto g_{\alpha\beta}$, we have a condition over the
ratios as follows:
\begin{equation}
\frac{R_{\left(  \alpha\beta\right)  }}{R_{s}}\varpropto\frac{g_{\alpha\beta}%
}{D} \tag{32}%
\end{equation}
that seems to be universal.

7)The equations are the simplest ones when $b^{-2}\ =0\ \left(  \beta
=0\right)  ,$taking the exact "quasilinear" form$\ \ \ $%
\begin{equation}
R_{\left(  \alpha\beta\right)  }-\frac{g_{\alpha\beta}}{4}R_{s}=\underset
{Maxwell-like}{\underbrace{\frac{R_{s}}{2\alpha^{2}}\left[  F_{\alpha\lambda
}F_{\beta}^{\ \lambda}-F_{\mu\nu}F^{\mu\nu}\frac{R_{\left(  \alpha
\beta\right)  }}{R_{s}}\right]  }}+2\lambda\underset{\widetilde{g}_{eff}%
}{\underbrace{\left[  g_{\alpha\beta}+\frac{1}{\mathcal{W}}F_{\alpha\lambda
}F_{\beta}^{\ \lambda}\right]  }}, \tag{33}%
\end{equation}
this particular case (e.g. projective invariant) will be used through this
work. Notice that when $b^{-2}\ =0\ \left(  \beta=0\right)  $ all terms into
the gravitational equation (31) involving the pseudoscalar invariant, namely
$F_{\mu\nu}\widetilde{F}^{\mu\nu}$ or $F_{\alpha\lambda}\widetilde{F}_{\beta
}^{\ \lambda}$ ,vanishes. Consequently we arrive to the simplest expression
(33) that will be used in Section XI\ for example.
\end{remark}

\subsection{$\delta_{f}L_{G}$}

Let us to take as starting point the geometrical Lagrangian (19)\
\begin{align}
\mathcal{L}_{gs}  &  =\sqrt{\det\left[  \lambda g_{\alpha\beta}\left(
1+\frac{R_{s}}{4\lambda}\right)  +\lambda F_{\alpha\beta}\left(  1+\frac
{R_{A}}{4\lambda}\right)  \right]  }\tag{34}\\
&  =\sqrt{\left\vert g\right\vert }\lambda^{2}\alpha^{2}\left(  \sqrt
{1+\frac{1}{2}\mathcal{F}_{\mu\nu}\mathcal{F}^{\mu\nu}-\frac{1}{16}\left(
\mathcal{F}_{\mu\nu}\widetilde{\mathcal{F}}^{\mu\nu}\right)  ^{2}}\right)
\tag{35}%
\end{align}
then, having into account that : $R_{A}=f^{\mu\nu}R_{\mu\nu}$ and
$\frac{\partial\ln\left(  \det F_{\mu\nu}\right)  }{\partial F_{\alpha\beta}%
}=f^{\alpha\beta}$ (due that $b$ that contains $R_{A}$ must be also included
in the variation)we obtain%

\begin{equation}
\text{\ \ }\frac{\delta L_{G}}{\delta F_{\sigma\omega}}=0\rightarrow\left(
\frac{\sqrt{\left\vert g\right\vert }\lambda\beta}{2\mathbb{R}b}\right)
\left[  \mathbb{F}^{\sigma\omega}\beta-\frac{\mathbb{F}}{4\lambda}R_{\left[
\mu\nu\right]  }\chi^{\mu\nu\sigma\omega}\right]  =0 \tag{36}%
\end{equation}
where: $\mathbb{F\equiv}\left[  F_{\mu\nu}F^{\mu\nu}-\frac{1}{4}b^{-2}\left(
F_{\mu\nu}\widetilde{F}^{\mu\nu}\right)  ^{2}\right]  $ , $\mathbb{F}%
^{\sigma\alpha}\mathbb{\equiv}\left[  F^{\sigma\alpha}-\frac{1}{4}%
b^{-2}\left(  F_{\mu\nu}\widetilde{F}^{\mu\nu}\right)  \widetilde{F}%
^{\sigma\alpha}\right]  $ and $\chi^{\mu\nu\sigma\omega}\equiv f^{\mu\omega
}f^{\sigma\nu}-f^{\mu\sigma}f^{\omega\nu}.$ Notice that the quantity
$b=\alpha/\beta$ (concretely $\beta)$ was also varied in the above expression
given the second term in (36).

Contracting (36) with $F_{\alpha\beta}$, a condition over the curvature and
the electromagnetic field invariants is obtained as
\[
\left(  \frac{\sqrt{\left\vert g\right\vert }\lambda\beta}{\mathbb{R}%
b}\right)  \mathbb{F}\left[  \beta-\frac{R_{A}}{2\lambda}\right]  =0
\]
This condition is satisfied for $R_{A}=-4\lambda$ is the exact projective
invariant case (that correspond with $\beta=0)$, and for $R_{A}=2\lambda.$

\begin{remark}
the variational equation (36) is a dynamic equation for the torsion field in
complete analogy with the eqs. (31) for the curvature.
\end{remark}

\section{Emergent trace free gravitational equations: the meaning of $\Lambda
$}

Starting from the trace free equation (31)\textit{ that is not assumed} but
arises from the model, the task\cite{tf} is to rewrite it as%

\begin{equation}
\underset{\equiv G_{\alpha\beta}}{\underbrace{\overset{\circ}{R}_{\alpha\beta
}-\frac{g_{\alpha\beta}}{2}\overset{\circ}{R}}}=\underset{\equiv
T_{\alpha\beta}^{h}}{\underbrace{6\left(  -h_{\alpha}h_{\beta}+\frac
{g_{\alpha\beta}}{2}h_{\gamma}h^{\gamma}\right)  }}+\frac{g_{\alpha\beta}}%
{2}R_{s}+T_{\alpha\beta}^{F}+2\lambda\rho_{\alpha\beta} \tag{37}%
\end{equation}
where%

\begin{align}
\rho_{\alpha\beta}  &  \equiv g_{\alpha\beta}+\frac{1}{\mathcal{W}}\left(
F_{\alpha\lambda}F_{\beta}^{\ \lambda}+\frac{F_{\mu\nu}\widetilde{F}^{\mu\nu}%
}{2b^{2}}\left(  \frac{F_{\eta\rho}\widetilde{F}^{\eta\rho}}{8}g_{\alpha\beta
}-F_{\alpha\lambda}\widetilde{F}_{\beta}^{\ \lambda}\right)  \right)
\tag{38}\\
T_{\alpha\beta}^{F}  &  \equiv\frac{R_{s}}{2\mathcal{W}}\left\{  \left(
F_{\alpha\lambda}F_{\beta}^{\ \lambda}-F_{\mu\nu}F^{\mu\nu}\frac{R_{\left(
\alpha\beta\right)  }}{R_{s}}\right)  +\right. \tag{39}\\
&  \left.  +\frac{1}{2b^{2}}\left[  F_{\mu\nu}\widetilde{F}^{\mu\nu}\left(
\frac{F_{\eta\rho}\widetilde{F}^{\eta\rho}}{8}g_{\alpha\beta}-F_{\alpha
\lambda}\widetilde{F}_{\beta}^{\ \lambda}\right)  +\frac{\left(  F_{\eta\rho
}\widetilde{F}^{\eta\rho}\right)  ^{2}}{2}\frac{R_{\left(  \alpha\beta\right)
}}{R_{s}}\right]  \right\} \nonumber
\end{align}
the LHS of (37) is the Einstein tensor. The "GR" divergence$\overset{\circ
}{\nabla}^{\alpha}$of $G_{\alpha\beta}$is zero because is a geometrical
geometrical identity and in an analog manner \ $\overset{\circ}{\nabla
}^{\alpha}\left(  T_{\alpha\beta}^{h}+T_{\alpha\beta}^{F}\right)  =0$ because
both tensors have the same symmetry that the corresponding GR\ energy momentum
tensors of a vector field and electromagnetic field respectively:
\[
\overset{\circ}{\nabla}^{\alpha}G_{\alpha\beta}=\overset{\circ}{\nabla
}^{\alpha}\left(  T_{\alpha\beta}^{h}+T_{\alpha\beta}^{F}\right)  =0
\]
consequently the remaining part must be a covariantly constant tensor that we
\textit{assume} proportional to $g_{\alpha\beta}$ :%
\[
\nabla^{\alpha}\left(  \frac{g_{\alpha\beta}}{2}R_{s}+2\lambda\rho
_{\alpha\beta}\right)  =0
\]%
\begin{equation}
\Rightarrow\left(  \frac{g_{\alpha\beta}}{2}R_{s}+2\lambda\rho_{\alpha\beta
}\right)  =\Lambda g_{\alpha\beta}\rightarrow R_{s}=2\Lambda\tag{40}%
\end{equation}
Coming back to the original trace free expressions we have the expected
formula%
\begin{equation}
\underset{\equiv G_{\alpha\beta}}{\underbrace{\overset{\circ}{R}_{\alpha\beta
}-\frac{g_{\alpha\beta}}{2}\overset{\circ}{R}}}=\underset{\equiv
T_{\alpha\beta}^{h}}{\underbrace{6\left(  -h_{\alpha}h_{\beta}+\frac
{g_{\alpha\beta}}{2}h_{\gamma}h^{\gamma}\right)  }}+T_{\alpha\beta}%
^{F}+\Lambda g_{\alpha\beta} \tag{41}%
\end{equation}

\begin{remark}
Tracing the first expression in (40) we have $R_{s}=2\Lambda=\overset{\circ
}{R}+6h_{\mu}h^{\mu}$ linking the value of the curvature and the norm of the
torsion vector field. Consequently, if the dual of the torsion field have the
role of the energy-matter carrier, the meaning of lambda as the vacuum energy
is immediately established.
\end{remark}

\begin{remark}
Notice that the LHS in expression (40) instead to be proportional to the
metric tensor it can be proportional to the square of a Killing-Yano tensor.
\end{remark}

\section{On the constancy of G}

At this level, no assertion can state with respect to $G$ or even with respect
to $c$. The link with the general relativistic case is given by the
identification of electromagnetic energy-momentum tensor with the term
analogous $T_{\alpha\beta}^{F}$ in our metric variational equations:
\[
\frac{8\pi G}{c^{4}}\left(  F_{\alpha\lambda}F_{\beta}^{\ \lambda}-F_{\mu\nu
}F^{\mu\nu}\frac{g_{\alpha\beta}}{4}\right)  \rightarrow\frac{R_{s}%
}{2\mathcal{W}}\left(  F_{\alpha\lambda}F_{\beta}^{\ \lambda}-F_{\mu\nu}%
F^{\mu\nu}\frac{R_{\left(  \alpha\beta\right)  }}{R_{s}}\right)
\]
Consequently we have:%

\begin{gather*}
\kappa=\frac{8\pi G}{c^{4}}\rightarrow\frac{R_{s}}{2\mathbb{R}^{2}\alpha^{2}%
}\\
and\text{ \ \ }\frac{g_{\alpha\beta}}{4}\text{ }=\frac{R_{\left(  \alpha
\beta\right)  }}{R_{s}}%
\end{gather*}
The above expression indicates that the ratio must remains constant due the
Noether symmetries and conservation laws of the field equations. Notice that
(as in the case of $b$) there exist a limit for all the physical fields coming
from the\textit{ geometrical invariants quantities.}

\section{The vector $h_{\mu}$ and the energy-matter interpretation}

One of the characteristics that more attract the attention in unified field
theoretical models is the possibility to introduce the energy and matter
through its geometrical structure. In our case the torsion field takes the
role of RHS\ of the standard GR\ gravity equation by mean its dual, namely
$h_{\mu}$.

Consequently, in order to explain the physical role of $h_{\mu},$ we know (due
the Hodge-de Rham decomposition [Appendix III]) that it can be decomposed as:%
\begin{equation}
h_{\alpha}=\nabla_{\alpha}\Omega+\varepsilon_{\alpha}^{\beta\gamma\delta
}\nabla_{\beta}A_{\gamma\delta}+\gamma_{1}\overset{axial\text{ }%
vector}{\overbrace{\varepsilon_{\alpha}^{\beta\gamma\delta}M_{\beta
\gamma\delta}}}+\gamma_{2}\overset{polar\text{ }vector}{\overbrace{P_{\alpha}%
}} \tag{42}%
\end{equation}
where $\gamma_{1}$and $\gamma_{2}$ can be phenomenologically related to
physical constants (e.g: $\gamma_{1}=\frac{8\pi}{c}\sqrt{G}$ is a physical
constant related to the Blackett formula \cite{bla}). The arguments in favour
of this type of theories and from the decomposition (42) can be resumed as follows:

i) the existence of an angular momentum Helmholtz theorem \cite{hel}%
\cite{hel1}: the theorem in analysis is exactly as in $E_{3}$ but, in the four
dimensional case $M_{4}$ there exists an additional \textit{axial vector};

iii) the concept of $chirality$ is achieved in the model by the existence of
polar and axial vectors in expression (42)$.$

iv) if $\Omega,A_{\gamma\delta}$ are the wave tensors and $\varepsilon
_{\alpha}^{\beta\gamma\delta}M_{\beta\gamma\delta},P_{\alpha}$ the particle
vectors (vector and axial part respectively), the concept of an inertial-wave
vector is introduced in the equation (42).

Consequently, from the eq.motion for the torsion namely: $\nabla_{\alpha
}T^{\alpha\beta\gamma}=-\lambda F^{\beta\gamma}$and coming back to (42) we
obtain the following important equation%
\begin{equation}
\overset{\circ}{\square}A_{\gamma\delta}-\gamma\left[  \nabla_{\alpha
}M_{\text{ \ }\gamma\delta}^{\alpha}+\left(  \nabla_{\gamma}P_{\delta}%
-\nabla_{\delta}P_{\gamma}\right)  \right]  =-\lambda F_{\gamma\delta}
\tag{43}%
\end{equation}
Let us consider, in particular, the case when $\lambda F_{\gamma\delta
}\rightarrow0:$%
\begin{equation}
\overset{\circ}{\square}A_{\gamma\delta}=\gamma\left[  \nabla_{\alpha
}M_{\text{ \ }\gamma\delta}^{\alpha}+\left(  \nabla_{\gamma}P_{\delta}%
-\nabla_{\delta}P_{\gamma}\right)  \right]  \tag{44}%
\end{equation}

We can immediately see that, if $M_{\text{ \ }\gamma\delta}^{\alpha}$ is
identified with the intrinsic spin angular momentum of the ponderable matter,
$P_{\delta}$ is its lineal momentum vector and $A_{\gamma\delta}$ is the
gravitational radiation tensor, then eq.(44) states that the sum of the
intrinsic spin angular momentum and the orbital angular momentum of ponderable
matter is conserved if the gravitational radiation is absent., if $M_{\text{
\ }\gamma\delta}^{\alpha}$ is identified with the intrinsic spin angular
momentum of the ponderable matter, $P_{\delta}$ is its lineal momentum vector
and $A_{\gamma\delta}$ is the gravitational radiation tensor, then eq.(44)
states that the sum of the intrinsic spin angular momentum and the orbital
angular momentum of ponderable matter is conserved if the gravitational
radiation is absent.

\subsection{Killing-Yano systems and the vector $h_{\mu}$}

Without enter in many details (these will be treated somewhere) the
antisymmetric tensor $A_{\gamma\delta}$ in the $h_{\beta}$ composition is
related with the Killing\cite{kil} and Killing-Yano\cite{kilyan} systems.
Consequently we can introduce two types of couplings into the $A_{\gamma
\delta}$ divergence : it correspond with the generalized current
interpretation that also has $h_{\mu}$.

i) Defining
\begin{equation}
A_{\gamma\delta}\equiv A_{\left[  \gamma;\delta\right]  } \tag{45}%
\end{equation}
such that
\begin{equation}
\overset{\circ}{\nabla}_{\rho}A_{\left[  \gamma;\delta\right]  }=\frac{4\pi
}{3}\left(  j_{\left[  \gamma\right.  }g_{\left.  \delta\right]  \rho}\right)
\tag{46}%
\end{equation}
then , in this case we can identify $A_{\gamma\delta}=2F_{\gamma\delta}$
because $F_{\gamma;\delta}^{\delta}=4j_{\gamma}$ and $A_{\left[  \gamma
\delta;\rho\right]  }=F_{\left[  \gamma\delta;\rho\right]  }=0$

In this case the contribution of $A_{\gamma\delta}$ to $h_{\beta}$ is null.

ii) Let us consider now a fully antisymmetric coupling as
\begin{equation}
A_{\left[  \gamma;\delta\right]  ;\rho}=\frac{4\pi}{3}j_{\left[
\gamma\right.  }F_{\left.  \delta\right]  \rho} \tag{47}%
\end{equation}
, having into account the vorticity vector also%
\begin{equation}
\omega_{\mu}\equiv u^{\lambda}\varepsilon_{\lambda\mu\nu\rho}\nabla^{\nu
}u^{\rho} \tag{48}%
\end{equation}
and considering a plasma with electrons, protons etc.
\begin{equation}
j^{\gamma}\sim A^{\gamma}+q_{s}n_{s}u_{s}^{\gamma} \tag{49}%
\end{equation}
where $A_{\mu}$ is the vector potential and $q_{s}$ is the particle charge,
$n_{s}$is the number density (in the rest frame) and the four-velocity of
species s is $u_{s}^{\gamma}$. In this case $h_{\alpha}$takes the form
\begin{align}
h_{\alpha}  &  =\nabla_{\alpha}\Omega+\varepsilon_{\alpha}^{\beta\gamma\delta
}\nabla_{\beta}A_{\gamma\delta}+\gamma_{1}\varepsilon_{\alpha}^{\beta
\gamma\delta}M_{\beta\gamma\delta}+\gamma_{2}P_{\alpha}\rightarrow\tag{50}\\
h_{\alpha}  &  =\nabla_{\alpha}\Omega+\varepsilon_{\alpha}^{\gamma\delta\rho
}\frac{4\pi}{3}j_{\left[  \gamma\right.  }F_{\left.  \delta\right]  \rho
}-\gamma_{1}u^{\lambda}\varepsilon_{\lambda\alpha\nu\rho}\nabla^{\nu}u^{\rho
}+\gamma_{2}P_{\alpha}\tag{51}\\
h_{\alpha}  &  =\nabla_{\alpha}\Omega+\varepsilon_{\alpha}^{\gamma\delta\rho
}\frac{4\pi}{3}\left[  A+q_{s}n_{s}u_{s}\right]  _{\left[  \gamma\right.
}F_{\left.  \delta\right]  \rho}-\gamma_{1}u^{\lambda}\varepsilon
_{\lambda\alpha\nu\rho}\nabla^{\nu}u^{\rho}+\gamma_{2}P_{\alpha} \tag{52}%
\end{align}
Consequently in 3+1 decomposition we have (overbar correspond to spacial
3-dim. vectors)
\begin{align}
h_{0}  &  =\nabla_{0}\Omega+\frac{4\pi}{3}\overline{j}\cdot\overline{B}%
+\gamma_{1}\overline{u}\cdot\left(  \overline{\nabla}\times\overline
{u}\right)  +\gamma_{2}P_{0}\tag{53}\\
h_{0}  &  =\nabla_{0}\Omega+\frac{4\pi}{3}\left[  \overline{A}\cdot\left(
\overline{\nabla}\times\overline{A}\right)  +q_{s}n_{s}\overline{u}_{s}%
\cdot\overline{B}\right]  +\gamma_{1}\overline{u}\cdot\left(  \overline
{\nabla}\times\overline{u}\right)  +\gamma_{2}P_{0} \tag{54}%
\end{align}
and
\begin{align}
h_{i}  &  =\nabla_{i}\Omega+\frac{4\pi}{3}\left[  -\left(  \overline{j}%
\times\overline{E}\right)  _{i}+j_{0}\overline{B}_{i}\right]  +\gamma
_{1}\left[  u_{0}\left(  \overline{\nabla}\times\overline{u}\right)  +\left(
\overline{u}\times\overline{\nabla}u_{0}\right)  +\left(  \overline{u}%
\times\overset{\cdot}{\overline{u}}\right)  \right]  _{i}+\gamma_{2}%
P_{i}\tag{55}\\
h_{i}  &  =\nabla_{i}\Omega+\frac{4\pi}{3}\left[  -\left(  \left(
\overline{A}+q_{s}n_{s}\overline{u}_{s}\right)  \times\overline{E}\right)
_{i}+\left(  \Phi+q_{s}n_{s}u_{0s}\right)  \overline{B}_{i}\right]
+\tag{56}\\
&  +\gamma_{1}\left[  u_{0}\left(  \overline{\nabla}\times\overline{u}\right)
+\left(  \overline{u}\times\overline{\nabla}u_{0}\right)  +\left(
\overline{u}\times\overset{\cdot}{\overline{u}}\right)  \right]  _{i}%
+\gamma_{2}P_{i}\nonumber
\end{align}
Notice that in $h_{0}$ we can recognize the magnetic and vortical helicities%
\begin{equation}
h_{0}=\nabla_{0}\Omega+\frac{4\pi}{3}\left[  h_{M}+q_{s}n_{s}\overline{u}%
_{s}\cdot\overline{B}\right]  +\gamma_{1}h_{V}+\gamma_{2}P_{0} \tag{57}%
\end{equation}
The above expression will be very important in the next sections, in
particular to discuss magnetogenesis and particle generation. Notice the
important fact that the symmetry of the vorticity can be associated to a
2-form bivector in the context of the notoph field\cite{ogpol}theory.

.

\section{Physical consequences}

In this Section we will make contact with the physical consequences of the
model. Firstly we introduce the 3+1 splitting for axisymmetric spacetimes that
is useful from the the physical viewpoint for the analysis of the
electrodynamic equations with high degree of nonlinearity, as in our case.
Secondly we take the 3+1 field equations in the in the linear limit where the
induction equations (dynamo) are obtained, showing explicitly the important
role of the torsion field as the generator of a purely geometric $\alpha
$-term. Thirdly, we derive the geometrical analog of the Lorentz force and the
elimination of the electric field from the induction equations. Also, the
origin of the seed magnetic field via the geometrical $\alpha$-term generated
by the torsion vector is worked out.

\subsection{Electrodynamic structure in 3+1}

The starting point will be the line element in $3+1$ splitting\cite{kt}%
\cite{kt1}(Appendix IV): the 4-dimensional space-time is split into
3-dimensional space and 1-dimensional time to form a foliation of
3-dimensional spacelike hypersurfaces. The metric of the space-time is
consequently, given by $ds^{2}=-\alpha^{2}dt^{2}+\gamma_{ij}\left(
dx^{i}+\beta^{i}dt\right)  \left(  dx^{j}+\beta^{j}dt\right)  $ where
$\gamma_{ij}$ is the metric of the 3-dimensional hypersurface,$\alpha$ is the
lapse function, and $\beta^{i}$ is the shift function (see Appendix IV for
details) . For any nonlinear Lagrangian, in sharp contrast with the
Einstein-Maxwell case, the field equations $d\ast\mathbb{F=\ast J}$ and the
Bianchi-geometrical condition $dF=0$ (where we have defined the Hodge dual
$\ast$ and $\mathbb{F=}\frac{\mathbb{\partial}\mathcal{L}}{\partial F})$ are
expressed by the vector fields%
\begin{equation}
E,B,\mathbb{E=}\frac{\mathbb{\partial}\mathcal{L}}{\partial E},\mathbb{B=}%
\frac{\mathbb{\partial}\mathcal{L}}{\partial B} \tag{58}%
\end{equation}
that live into the slice. In our case given by the geometrical Lagrangian
$\mathcal{L}_{g}$ (not be confused with the Lie derivative$\mathcal{L}%
_{\mathcal{\beta}}!)$
\begin{align}
\nabla\cdot\mathbb{E}  &  =-\overline{h}\cdot\mathbb{B}+4\pi\rho_{e}\tag{59}\\
\nabla\cdot B  &  =0\tag{60}\\
\nabla\times\left(  \alpha E\right)   &  =-(\partial_{t}-\mathcal{L}%
_{\mathcal{\beta}})B\nonumber\\
&  =-\partial_{0}B+\left(  \beta\cdot\nabla\right)  B-\left(  B\cdot
\nabla\right)  \beta\tag{61}\\
\nabla\times\left(  \alpha\mathbb{B}\right)  +h_{0}\mathbb{B-}\overline
{h}\times\mathbb{E}  &  =-(\partial_{t}-\mathcal{L}_{\mathcal{\beta}%
})\mathbb{E}+4\pi\alpha j\nonumber\\
&  =\partial_{0}\mathbb{E}-\left(  \beta\cdot\nabla\right)  \mathbb{E}+\left(
\mathbb{E}\cdot\nabla\right)  \beta+4\pi\alpha j \tag{62}%
\end{align}
where $h^{\mu}$ is the torsion vector. Notice that, here and the subsequent
Sections, the overbar indicates 3-dimensional space vectors.

\subsection{Dynamo effect and geometrical origin of $\alpha\Omega$ term}

In the case of weak field approximation and $\left(  F^{01}\rightarrow
E^{i},\text{ }F^{jk}\rightarrow B^{i}\right)  $the electromagnetic
Maxwell-type equations in 3+1 take the form
\begin{equation}
\nabla_{\nu}F^{\nu\mu}=T^{\mu\nu\rho}F_{\nu\rho}=\varepsilon_{\text{
\ \ \ \ \ }\delta}^{\mu\nu\rho}h^{\delta}F_{\nu\rho}\text{ \ \ \ \ \ \ }%
\left(  d^{\ast}F=^{\ast}J\right)  \tag{63}%
\end{equation}%
\begin{equation}
\overline{\nabla}\cdot\overline{E}=-\overline{h}\cdot\overline{B} \tag{64}%
\end{equation}%
\begin{equation}
\partial_{t}\overline{E}-\overline{\nabla}\times\overline{B}=h^{0}\overline
{B}-\overline{h}\times\overline{E} \tag{65}%
\end{equation}
and
\begin{equation}
\nabla_{\nu}^{\text{ \ }\ast}F^{\nu\mu}=0\text{\ \ \ \ \ \ }\left(
dF=0\right)  \tag{66}%
\end{equation}%
\begin{equation}
\overline{\nabla}\cdot\overline{B}=0 \tag{67}%
\end{equation}%
\begin{equation}
\partial_{t}\overline{B}=-\overline{\nabla}\times\overline{E} \tag{68}%
\end{equation}
Putting all together, the set of equations is
\begin{align}
\overline{\nabla}\cdot\overline{E}+\overline{h}\cdot\overline{B}  &
=\rho_{ext}\tag{69}\\
\partial_{t}\overline{E}-\overline{\nabla}\times\overline{B}  &
=h^{0}\overline{B}-\overline{h}\times\overline{E}-\sigma_{ext}\left[
\overline{E}+\overline{v}\times\overline{B}\right] \tag{70}\\
\overline{\nabla}\cdot\overline{B}  &  =0\tag{71}\\
\partial_{t}\overline{B}  &  =-\overline{\nabla}\times\overline{E} \tag{72}%
\end{align}
where we have introduced external charge density and current. Following the
standard procedure we take the rotational to the second equation above
obtaining straightforwardly the modified dynamo equation%
\begin{equation}
\overline{\nabla}\times\partial_{t}\overline{E}+\overline{\nabla}^{2}%
\overline{B}=\overline{\nabla}\times\left(  h^{0}\overline{B}\right)  +\left(
\overline{h}\cdot\overline{B}-\rho_{ext}\right)  \overline{h}+\left(
\overline{\nabla}\cdot\overline{h}\right)  \overline{E}-\sigma_{ext}\left[
\partial_{t}\overline{B}+\left(  \overline{\nabla}\cdot\overline{v}\right)
\overline{B}\right]  \tag{73}%
\end{equation}
where the standard identities of the vector calculus plus the first, the third
and the fourth equations above have been introduced. Notice that in the case
of the standard approximation and (in the spirit of this research) without any
external or additional ingredients, we have%
\begin{equation}
\overline{\nabla}^{2}\overline{B}=h^{0}\left(  \overline{\nabla}%
\times\overline{B}\right)  +\left(  \overline{h}\cdot\overline{B}\right)
\overline{h}+\left(  \overline{\nabla}\cdot\overline{h}\right)  \overline{E}
\tag{74}%
\end{equation}
Here we can see that there exist and $\alpha$- term with a pure geometrical
origin (and not only a turbulent one) that is given by $h^{0}$ (the zero
component of the dual of the torsion tensor).

\subsection{The generalized Lorentz force}

An important point in any theory beyond relativity is the concept of force. As
is known, general relativity has deficiencies at this point. Now we are going
to show that it is possible to derive from our proposal the Lorentz force as
follows. From expression (32) the geometrical induced current is recognized
\begin{equation}
\partial_{t}\overline{E}-\overline{\nabla}\times\overline{B}=h^{0}\overline
{B}-\overline{h}\times\overline{E}\equiv\overline{J} \tag{77}%
\end{equation}

\begin{align}
\overline{J}\times\overline{B}  &  =\left(  h^{0}\overline{B}-\overline
{h}\times\overline{E}-\overline{j}_{ext}\right)  \times\overline{B}\tag{78}\\
&  =-\left[  \left(  \overline{h}\cdot\overline{B}\right)  \overline
{E}-\left(  \overline{E}\cdot\overline{B}\right)  \overline{h}\right]
-\overline{j}_{ext}\times\overline{B} \tag{79}%
\end{align}
we assume $j_{ext}$ proportional to the velocity and other contributions.
Consequently, reordering terms from above, a geometrically induced
Lorentz-like force arises%
\begin{align}
\left(  \overline{J}+\overline{j}_{ext}\right)  \times\overline{B}  &
=-\left[  \underset{\rho_{geom}}{\underbrace{\left(  \overline{h}%
\cdot\overline{B}\right)  }}\overline{E}-\left(  \overline{E}\cdot\overline
{B}\right)  \overline{h}\right]  \rightarrow\tag{80}\\
\underset{\rho_{geom}}{\underbrace{\left(  \overline{h}\cdot\overline
{B}\right)  }}\overline{E}+\underset{j_{gen}}{\underbrace{\left(  \overline
{J}+\overline{j}_{ext}\right)  }}\times\overline{B}  &  =\left(  \overline
{E}\cdot\overline{B}\right)  \overline{h}\rightarrow\text{Lorentz induced
force} \tag{81}%
\end{align}
being the responsible of the induced force, the torsion vector itself. Notice,
from the above equation, the following issues:

1)The external currents are identified with $\overline{J}$

\bigskip

2) We can eliminate the electric field in standard form%
\begin{equation}
\overline{E}=\frac{\left(  \overline{E}\cdot\overline{B}\right)  \overline
{h}-\overline{j}_{gen}\times\overline{B}}{\left(  \overline{h}\cdot
\overline{B}\right)  } \tag{82}%
\end{equation}
being the above expression very important in order to replace the electric
field into the dynamo equation, introducing naturally the external current in
the model.

\subsection{Generalized current and $\alpha$-term}

In previous paragraph we have derived a geometrical induced Lorentz force
where the link between the physical world and the proposed geometrical model
is through a generalized current $j_{gen}.$ An important fact of that
expression is that it is possible to eliminate the electric field (and insert
it into the equation of induction) as follows.

From the formula of the induction, namely%

\begin{equation}
\overline{\nabla}^{2}\overline{B}\underset{\mathcal{E}_{Geom}}{\underbrace
{+\overline{\nabla}\times\left(  -h^{0}\overline{B}+\overline{h}%
\times\overline{E}\right)  }}=0 \tag{83}%
\end{equation}
and using the eq. (49)\ to eliminate the electric field as function of the
torsion, the generalized current and the magnetic field respectively:
\begin{align}
\overline{h}\times\overline{E}  &  =\frac{-\overline{h}\times\left(
\overline{j}_{gen}\times\overline{B}\right)  }{\left(  \overline{h}%
\cdot\overline{B}\right)  }=-\frac{\left(  \overline{h}\cdot\overline
{B}\right)  \overline{j}_{gen}-\left(  \overline{h}\cdot\overline{j}%
_{gen}\right)  \overline{B}}{\left(  \overline{h}\cdot\overline{B}\right)
}\tag{84}\\
\overline{h}\times\overline{E}  &  =-\overline{j}_{gen}+\frac{\left(
\overline{h}\cdot\overline{j}_{gen}\right)  }{\left(  \overline{h}%
\cdot\overline{B}\right)  }\overline{B}=\left\vert \overline{j}_{gen}%
\right\vert \left(  -n_{\overline{j}_{gen}}+\frac{\cos\alpha}{\cos\beta}%
n_{B}\right)  \tag{85}%
\end{align}
being $\alpha$ the angle between the vector torsion $\overline{h}$ and the
generalized current $\overline{j}_{gen}$and $\beta$ the angle between
$\overline{h}$ and the magnetic field $\overline{B}.$Above, $n_{B}$ and
$n_{\overline{j}_{gen}}$ are unitary vectors in the direction of $\overline
{B}$ and $\overline{j}_{gen}$ respectively. Notice the important fact that the
RHS\ of (68) is independent of \ the torsion and the magnetic field.
Consequently we obtain%
\begin{equation}
\overline{\nabla}^{2}\overline{B}\underset{\mathcal{E}_{Geom}}{\underbrace
{+\overline{\nabla}\times\left[  -\overline{j}_{gen}+\left(  \frac{\left(
\overline{h}\cdot\overline{j}_{gen}\right)  }{\left(  \overline{h}%
\cdot\overline{B}\right)  }-h^{0}\right)  \overline{B}\right]  }}=0 \tag{86}%
\end{equation}
We introduce the explicitly the physical scenario via the generalized current
$\overline{j}_{gen}$%
\begin{equation}
-\overline{j}_{gen}\sim\sigma_{ext}\left[  \overline{E}+\overline{v}%
\times\overline{B}\right]  +\left(  \frac{c}{e}\frac{\overline{\nabla}p}%
{n_{e}}\right)  \tag{87}%
\end{equation}
then%
\begin{equation}
\overline{\nabla}^{2}\overline{B}\underset{\mathcal{E}_{Geom}}{\underbrace
{+\overline{\nabla}\times\left[  \sigma_{ext}\left[  \overline{E}+\overline
{v}\times\overline{B}\right]  +\left(  \frac{c}{e}\frac{\overline{\nabla}%
p}{n_{e}}\right)  +\left(  \frac{\left(  \overline{h}\cdot\overline{j}%
_{gen}\right)  }{\left(  \overline{h}\cdot\overline{B}\right)  }-h^{0}\right)
\overline{B}\right]  }}=0 \tag{88}%
\end{equation}%
\begin{equation}
\overline{\nabla}^{2}\overline{B}\underset{\mathcal{E}_{Geom}}{\underbrace
{+\sigma_{ext}\left[  \left(  -\partial_{t}\overline{B}\right)  +\overline
{\nabla}\times\left(  \overline{v}\times\overline{B}\right)  \right]
+\overline{\nabla}\times\left(  \frac{c}{e}\frac{\overline{\nabla}p}{n_{e}%
}\right)  +\overline{\nabla}\times\left(  \frac{\left(  \overline{h}%
\cdot\overline{j}_{gen}\right)  }{\left(  \overline{h}\cdot\overline
{B}\right)  }-h^{0}\right)  \overline{B}}}=0 \tag{89}%
\end{equation}
finally the expected geometrically induced expression is obtained:%
\begin{equation}
\partial_{t}\overline{B}=\eta\overline{\nabla}^{2}\overline{B}+\overline
{\nabla}\times\left(  \overline{v}\times\overline{B}\right)  +\eta
\overline{\nabla}\times\left[  \left(  \frac{c}{e}\frac{\overline{\nabla}%
p}{n_{e}}\right)  +\alpha\overline{B}\right]  =\partial_{t}\overline{B}
\tag{90}%
\end{equation}%
\begin{equation}
\rightarrow\eta\underset{diffusive}{\underbrace{\overline{\nabla}^{2}%
\overline{B}}}+\underset{advective}{\underbrace{\overline{\nabla}\times\left(
\overline{v}\times\overline{B}\right)  }}+\eta\underset{\alpha-term}%
{\underbrace{\overline{\nabla}\times\left(  \alpha\overline{B}\right)  }%
}+\underset{Biermann\text{ }battery}{\underbrace{\frac{c}{e}\frac
{\overline{\nabla}p\times\overline{\nabla}n_{e}}{n_{e}^{2}}}}=-\partial
_{t}\overline{B} \tag{91}%
\end{equation}
where $\eta\equiv\frac{1}{\sigma_{ext}}$ as usual and the geometric $\alpha$:
\begin{align}
\alpha &  \equiv\left(  \frac{\left(  \overline{h}\cdot\overline{j}%
_{gen}\right)  }{\left(  \overline{h}\cdot\overline{B}\right)  }-h^{0}\right)
\tag{92}\\
&  =\left(  \frac{\cos\alpha\left\vert \overline{j}_{gen}\right\vert }%
{\cos\beta\left\vert \overline{B}\right\vert }-h^{0}\right) \nonumber
\end{align}

\subsection{Seed magnetic field}

Notice from the last expression that $\alpha\overline{B}$ is explicitly
\begin{equation}
\alpha\overline{B}=\frac{\cos\alpha\left\vert \overline{j}_{gen}\right\vert
}{\cos\beta}n_{B}-h^{0}\overline{B} \tag{93}%
\end{equation}
or (via elimination of the unitary vector)%
\begin{equation}
\alpha\left\vert \overline{B}\right\vert =\frac{\cos\alpha\left\vert
\overline{j}_{gen}\right\vert }{\cos\beta}-h^{0}\left\vert \overline
{B}\right\vert \tag{94}%
\end{equation}
we see \textit{clearly} the first term in RHS independent of the intensity of
the magnetic field. Considering only the terms of interest without the
diffusive and advective term in the induction equation(only time-dependence
for the magnetic field is preserved) namely%
\begin{align}
\eta\underset{\alpha-term}{\underbrace{\overline{\nabla}\times\left(
\alpha\overline{B}\right)  }}  &  =-\partial_{t}\overline{B}\tag{96}\\
\eta\overline{\nabla}\left(  \frac{\cos\alpha\left\vert \overline{j}%
_{gen}\right\vert }{\cos\beta}\right)   &  =-\partial_{t}\left\vert
\overline{B}\right\vert \tag{97}%
\end{align}
we see that the currents given by the fields (related to the geometry via
$h_{\alpha})$ originate the magnetic field.

If we consider all the currents of the fields of theory (fermions, bosons,
etc.) the seed would be precisely these field currents. The other missing
point is to derive the fluid (hydrodynamic) equations (which as is known does
not have a definite Lagrangian formulation) from the same unified formulation.
Notice that there are, under special conditions, analogous formulas for
vorticity $\omega$ than for the magnetic field $B$. This would mean that the
2-form of vorticity must also be included in the fundamental antisymmetric
tensor, together with the electromagnetic field.

\subsection{Comparison with the mean field formalism}

Now we compare the obtained equations with respect to the mean field
formalism\cite{r}. Starting from expressions (69-72) as before, we have:%
\begin{equation}
\eta\overline{\nabla}^{2}\overline{B}+\overline{\nabla}\times\left(
\overline{v}\times\overline{B}\right)  -\partial_{t}\overline{B}%
\underset{\mathcal{E}_{Geom}}{\underbrace{+\eta\overline{\nabla}\times\left(
-h^{0}\overline{B}+\overline{h}\times\overline{E}\right)  }}=0 \tag{98}%
\end{equation}

$\mathcal{E}_{Geom}$ takes the place of electromotive force due the torsion
field with full analogy as $\mathcal{E=}\left\langle u\times b\right\rangle $
is the mean electromotive force due to fluctuations. Also as in the mean field
case that there are the splitting%
\begin{equation}
\mathcal{E}=\mathcal{E}^{\left\langle 0\right\rangle }+\mathcal{E}%
^{\left\langle \overline{B}\right\rangle } \tag{99}%
\end{equation}
with $\mathcal{E}^{\left\langle 0\right\rangle }$ independent of $\left\langle
\overline{B}\right\rangle $ and $\mathcal{E}^{\left\langle \overline
{B}\right\rangle }$linear and homogeneous in $\overline{B}$, we have in the
torsion case the following correspondence%
\begin{align*}
-h^{0}\overline{B}  &  \longleftrightarrow\mathcal{E}^{\left\langle
\overline{B}\right\rangle }\\
\overline{h}\times\overline{E}  &  \longleftrightarrow\mathcal{E}%
^{\left\langle 0\right\rangle }\\
geometrical  &  \longleftrightarrow turbulent
\end{align*}
Consequently, the problems of mean-field dynamo theory that are concerned with
the generation of a mean EMF by turbulence, have in this model a pure
geometric counterpart. In the past years, attention has shifted from kinematic
calculations, akin to those familiar from quasilinear theory for plasmas, to
self-consistent theories which account for the effects of small scale magnetic
fields (including their back-reaction on the dynamics) and for the constraints
imposed by the topological conservation laws, such as that for magnetic
helicity. Here the torsion vector generalize (as we can see from above set of
equations) the concept of helicity. The consequence of this role of the dual
torsion field is that the traditionally invoked mean-field dynamo mechanism
(i.e. the so-called alpha effect) may be severely quenched or increased at
modest fields and magnetic Reynolds numbers, and that spatial transport of
this generalized magnetic helicity is crucial to mitigating this quench. Thus,
the dynamo problem is seen in our model as one of generalized helicity
transport, and so may be tackled like other problems in turbulent transport. A
key element in this approach is to understand the evolution of the torsion
vector field besides of the turbulence energy and the generalized helicity
profiles in space-time. This forces us to confront the problem of spreading of
strong MHD turbulence, and a spatial variant or analogue of the selective
decay problem with the dynamics of the torsion field.

\section{Torsion, axion electrodynamics vs. Chern Simons theory}

Let us review briefly the electromagnetic sector of the theory QCD based in a
gauge symmetry $SU\left(  3\right)  \times U\left(  1\right)  $%
\begin{gather}
L_{QCD/QED}=+\sum\overline{\psi}_{f}\left[  \gamma^{\mu}\left(  \partial_{\mu
}-ig_{f}t^{\alpha}A_{\mu}^{\alpha}-iq_{f}A_{\mu}\right)  -m_{f}\right]
\psi_{f}-\tag{100}\\
-\frac{G_{\mu\nu}^{\alpha}G^{\alpha\mu\nu}}{4}-\frac{F_{\mu\nu}F^{\mu\nu}}%
{4}-\frac{g^{2}\theta G_{\mu\nu}^{\alpha}\widetilde{G}^{\alpha\mu\nu}}%
{32\pi^{2}}-\frac{g^{2}\theta F_{\mu\nu}\widetilde{F}^{\mu\nu}}{32\pi^{2}%
},\nonumber
\end{gather}
As is well know, electromagnetic fields will couple to the electromagnetic
currents, namely:$J_{\mu}=\underset{f}{\sum}q_{f}\overline{\psi}_{f}%
\gamma_{\mu}\psi_{f}$ \ consequently , there appear term will induce through
the quark loop the coupling of $F_{\mu\nu}\widetilde{F}^{\mu\nu}$ \ (the
anomaly) to the QCD topological charge . The effective Lagrangian can be
written as
\begin{equation}
L_{MCS}=-\frac{F_{\mu\nu}F^{\mu\nu}}{4}-A_{\mu}J^{\mu}-\frac{c}{4}\theta
F_{\mu\nu}\widetilde{F}^{\mu\nu} \tag{101}%
\end{equation}
where a pseudo-scalar field $\theta=\theta(\overline{x},t)$ (playing the role
of the axion field) is introduced and $c=\underset{f}{\sum}\frac{\left(
q_{f}e\right)  ^{2}}{2\pi^{2}}$ . This is the Chern-Simons Lagrangian where,
if $\theta$is constant, the last term is a total divergence: $F_{\mu\nu
}\widetilde{F}^{\mu\nu}=\partial_{\mu}J_{CS}^{\mu}.$The question appear if
$\theta$ is not a constant $\theta F_{\mu\nu}\widetilde{F}^{\mu\nu}%
=\theta\partial_{\mu}J_{CS}^{\mu}=\partial_{\mu}\left(  \theta J_{CS}^{\mu
}\right)  -J_{CS}^{\mu}\partial_{\mu}\theta$

Now we can see from the previous section that if, from the general
decomposition of the four dimensional dual of the torsion field via the Hodge
de Rham theorem we retain $b_{\alpha}$ as gradient of a pseudoscalar (e.g:
axion) these equations coincide in form with the respective equation for
MCS\ theory. Precisely because under this condition $h_{\alpha}=\nabla
_{\alpha}\theta$ , in flat space (curvature=0 but torsion$\neq0)$ the
equations become the same as in\cite{dk} namely%
\begin{align}
\overline{\nabla}\cdot\overline{E}-c\overline{P}\cdot\overline{B}  &
=\rho_{ext}\tag{102}\\
\partial_{t}\overline{E}-\overline{\nabla}\times\overline{B}  &
=-c\overset{\cdot}{\theta}\overline{B}+c\overline{P}\times\overline{E}%
-\sigma_{ext}\left[  \overline{E}+\overline{v}\times\overline{B}\right]
\tag{103}\\
\overline{\nabla}\cdot\overline{B}  &  =0\tag{104}\\
\partial_{t}\overline{B}  &  =-\overline{\nabla}\times\overline{E} \tag{105}%
\end{align}
provided:%
\begin{align}
h^{0}  &  \rightarrow-c\overset{\cdot}{\theta}\tag{106}\\
\overline{h}  &  \rightarrow-c\overline{P} \tag{107}%
\end{align}
where from QCD\ the constant $c$ is determined as $c=\frac{e^{2}}{2\pi}$ and
the $\partial_{\mu}\theta=\left(  \overset{\cdot}{\theta},\overline{P}\right)
$ in the\cite{dk} notation. The main difference is that while in the case of
photons in axion ED was given by\cite{wil} the Lagrangian where that above
equations are derived is
\begin{equation}
L_{MCS}=-\frac{F_{\mu\nu}F^{\mu\nu}}{4}-A_{\mu}J^{\mu}+\frac{c}{4}P_{\mu
}J_{CS}^{\mu},\text{ \ \ \ \ \ \ }J_{CS}^{\mu}\equiv\varepsilon^{\mu\sigma
\rho\nu}\text{\ }A_{\sigma}F_{\rho\nu} \tag{108}%
\end{equation}
in our case is the dual of the torsion field (that we take as the gradient of
a pseudoscalar) responsible of the particular structure of the set of equations.

\section{Magnetic helicity generation and cosmic torsion field}

Here we consider the projective invariant case: $\beta=0$ $\left(
R_{A}=-4\lambda\right)  $where the gravitational and field equations are
considerably simplified because $\mathbb{R}=1$ and $b^{-1}=0$ . Scalar
curvature $R$ and the torsion 2-form field $T_{\mu\nu}^{a}$ with a $SU\left(
2\right)  -$Yang-Mills structure are defined in terms of the affine connection
$\Gamma_{\mu\nu\text{ }}^{\lambda}$ and the $SU(2)$ valuated (structural
torsion potential) $f_{\ \mu\text{ }}^{a}$by
\begin{equation}
R=g^{\mu\nu}R_{\mu\nu}\hspace{1cm}R_{\mu\nu}=R_{\mu\lambda\nu}^{\lambda}
\tag{109}\label{eq38}%
\end{equation}%
\[
R_{\mu\lambda\nu}^{\lambda}=\partial_{\nu}\Gamma_{\mu\rho\text{ }}^{\lambda
}-\partial_{\rho}\Gamma_{\mu\nu\text{ }}^{\lambda}+...
\]%
\[
T_{\ \mu\nu}^{a}=\partial_{\mu}f_{\ \nu\text{ }}^{a}-\partial_{\nu}%
f_{\ \mu\text{ }}^{a}+\varepsilon_{bc}^{a}f_{\ \mu\text{ }}^{b}f_{\ \nu\text{
}}^{c}%
\]
$G$ and $\Lambda$ are the geometrically induced Newton gravitational constant
(as we have been discussed before) and the integration cosmological constant,
respectively. From the last equation for the totally antisymmetric Torsion
2-form, the potential $f_{\ \mu\text{ }}^{a}$ define the affine connection
$\Gamma_{\mu\nu\text{ }}^{\lambda}$.Similarly to the case of
Einstein-Yang-Mills systems, for our new $UFT$\ model it can be interpreted as
a prototype of gauge theories interacting with gravity (\emph{e.g.} QCD, GUTs,
etc.). We stress here the important fact that all the fundamental constants
are really geometrically induced as required by the Mach principle. After
varying the action, we obtain the gravitational equation (41), namely
\begin{equation}
\overset{\circ}{R}_{\alpha\beta}-\frac{g_{\alpha\beta}}{2}\overset{\circ}%
{R}=6\left(  -h_{\alpha}h_{\beta}+\frac{g_{\alpha\beta}}{2}h_{\gamma}%
h^{\gamma}\right)  +\kappa_{geom}\left[  F_{\alpha\lambda}F_{\beta}%
^{\ \lambda}-F_{\mu\nu}F^{\mu\nu}\frac{g_{\alpha\beta}}{4}\text{ }\right]
+\Lambda g_{\alpha\beta} \tag{110}\label{eq39}%
\end{equation}
with the "gravitational constant" geometrically induced as%
\begin{equation}
\kappa_{g}\equiv\frac{R_{s}}{2\mathcal{W}}=\left.  \frac{8\pi G}{c^{4}%
}\right\vert _{today} \tag{111}\label{cons}%
\end{equation}
and the field equation for the torsion 2-form in differential form
\begin{equation}%
\begin{array}
[c]{rcl}%
d^{\ast}T^{a}+\frac{1}{2}\varepsilon^{abc}\left(  f_{b}\wedge^{\ast}%
T_{c}-^{\ast}T_{b}\wedge f_{c}\right)  & = & -\lambda\mathbb{\ }^{\ast}f^{a}%
\end{array}
\tag{112}\label{eq40}%
\end{equation}

Notice that\ref{cons} $\kappa_{g}$ and $\Lambda$ are not independent, but
related by \ $R_{s}=2\Lambda$. In this case $\beta=0$ we have the simplest
expression:%
\[
\kappa_{g}\equiv\frac{R_{s}}{2\left(  1+\frac{R_{s}}{4\lambda}\right)  ^{2}%
}=\frac{\Lambda}{\left(  1+\frac{2\Lambda}{4\lambda}\right)  ^{2}}%
\]
in consequence, generalizing the conjecture of Markov in\cite{mar}\textbf{, if
}$\Lambda$\textbf{ is proportional to the energy, }$\kappa$ goes as\textbf{
}$\Lambda$ if $\left\vert \Lambda\right\vert \leq1,$ and as $\Lambda^{-1}$ in
other case.

We are going to seek for a classical solution of (\ref{eq39}) and (\ref{eq40})
with the following ansatz for the metric and gauge connection
\begin{equation}
ds^{2}=d\tau^{2}+a^{2}\left(  \tau\right)  \sigma^{i}\otimes\sigma^{i}\equiv
d\tau^{2}+e^{i}\otimes e^{i}. \tag{113}\label{eq41}%
\end{equation}
Here $\tau$ is the euclidean time and the dreibein is defined by $e^{i}\equiv
a\left(  \tau\right)  \sigma^{i}$. The gauge connection is
\begin{equation}
f^{a}\equiv f_{\mu}^{a}dx^{\mu}=f\sigma^{a}, \tag{114}\label{eq42}%
\end{equation}
for $a,b,c=1,2,3$, and for $a,b,c=0$ we have
\begin{equation}
f^{0}\equiv f_{\mu}^{0}dx^{\mu}=s\sigma^{0}. \tag{115}\label{eq43}%
\end{equation}
This choice for the potential torsion is accordingly to the symmetries
involved in the problem.

The $\sigma^{i}$ 1-form satisfies the $SU\left(  2\right)  $ Maurer-Cartan
structure equation
\begin{equation}
d\sigma^{a}+\varepsilon_{\ bc}^{a}\sigma^{b}\wedge\sigma^{c}=0 \tag{116}%
\label{eq44}%
\end{equation}
Notice that in the ansatz the frame and $SU\left(  2\right)  $ (isospin-like)
indices are identified (as for the case with the non-abelian-Born-Infeld (NBI)
Lagrangian of ref. \cite{ref3}) The torsion 2-form
\begin{equation}
T^{\gamma}=\frac{1}{2}T_{\ \mu\nu}^{\gamma}dx^{\mu}\wedge dx^{\nu}
\tag{117}\label{eq45}%
\end{equation}
becomes
\begin{align}
T^{a}  &  =df^{a}+\frac{1}{2}\varepsilon_{\ bc}^{a}f^{b}\wedge f^{c}%
\tag{118}\label{eq46}\\
&  =\left(  -f+\frac{1}{2}f^{2}\right)  \varepsilon_{\ bc}^{a}\sigma^{b}%
\wedge\sigma^{c}\nonumber
\end{align}%
\begin{equation}%
\begin{array}
[c]{rcl}%
d^{\ast}T^{a}+\frac{1}{2}\varepsilon^{abc}\left(  f_{b}\wedge^{\ast}%
T_{c}-^{\ast}T_{b}\wedge f_{c}\right)  & = & -2\lambda\mathbb{\ }^{\ast}%
f^{a}\\
(-2f+f^{2})(1-f)d\tau\wedge e^{b}\wedge e^{c} & = & -2\lambda d\tau\wedge
e^{b}\wedge e^{c}%
\end{array}
\tag{119}\label{eq59}%
\end{equation}%
\begin{equation}
^{\ast}T^{a}\mathbb{\equiv}h(-2f+f^{2})d\tau\wedge\frac{e^{a}}{a^{2}}
\tag{120}\label{eq60}%
\end{equation}%
\begin{equation}
^{\ast}f^{a}=-f\frac{d\tau\wedge e^{b}\wedge e^{c}}{a^{3}} \tag{121}%
\label{eq61}%
\end{equation}
Note that to be complete in our description of the possible physical
scenarios, we include $f^{0}$ as an $U\left(  1\right)  $ component of the
torsion potential (although does not belong to the space $SU(2)/U(1)$). Having
all the above issues into account, the expression for the torsion is analogous
to the non-abelian 2-form strength field of \cite{ref3}.

Inserting $T^{a}$ from (\ref{eq46}) into the dynamic equation (\ref{eq40}) we
obtain
\begin{equation}
(-2f+f^{2})(1-f)d\tau\wedge e^{b}\wedge e^{c}=-\lambda d\tau\wedge e^{b}\wedge
e^{c}, \tag{122}\label{eq47}%
\end{equation}
and from expression (\ref{eq47}) we have an algebraic cubic equation for $f$%
\begin{equation}
(-2f+f^{2})(1-f)+\lambda=0 \tag{123}\label{eq52}%
\end{equation}

We can see that, in contrast with our previous work with a dualistic theory
\cite{ref3}where the NBI energy-momentum tensor of Born-Infeld was considered,
there exist three non trivial solutions for $f$, depending on the cosmological
constant $\lambda.$ In this preliminary analysis of the problem, only the
values of $f$ that make the quantity $\left(  -f+\frac{1}{2}f^{2}\right)  $
$\in\mathbb{R}$. Consequently for $\lambda=2$ we find $f=2.35$ then
\begin{equation}
T_{bc}^{a}=\frac{2}{5}\frac{\varepsilon_{bc}^{a}}{a^{2}};\ \ \ \ \ \ \ \hspace
{1.54cm}T_{0c}^{a}=0 \tag{124}\label{eq53}%
\end{equation}
That is, only spatial torsion field is non vanishing while cosmic time torsion
field vanishes(an analogous feature with magnetic and electric Yang-Mills can
be seen in the solution of Giddings and Strominger and in \cite{ref3}).
Substituting the expression for the torsion 2-form (\ref{eq53}) $^{1}%
$\footnotetext[1]{in the tetrad: $\overset{\circ}{R}_{_{00}}=-3\frac
{\overset{\cdot\cdot}{a}}{a},\overset{\circ}{R}_{ab}=-\left[  \frac
{\overset{\cdot\cdot}{a}}{a}+2\left(  \frac{\overset{\cdot}{a}}{a}\right)
^{2}-\frac{2}{a^{2}}\right]  $} into the symmetric part of the variational
equation we reduce the gravitational equations to an ordinary differential
equation for the scale factor $a$,%
\begin{equation}
3\left[  \left(  \frac{\overset{.}{a}}{a}\right)  ^{2}-\frac{1}{a^{2}}\right]
-\Lambda=\frac{3\kappa_{g}}{4a^{2}}\left(  f^{2}+s^{2}\right)  +\frac
{3}{2a^{4}}f^{2}\left(  f-2\right)  ^{2} \tag{125}\label{eq56}%
\end{equation}
that in the case for the computed value for $f\sim2.35$ with $s=10$ and
$\Lambda\lesssim1$ the scale factor is described in Figure 1 and the scale
factor goes as:%
\begin{equation}
a\left(  \tau\right)  =\Lambda^{-1/2}\sqrt{\left(  1-\frac{12\kappa_{g}%
^{2}\Lambda}{\alpha}\right)  ^{1/2}\sinh\left(  \sqrt{\Lambda/3}\left(
\tau-\tau_{0}\right)  \right)  -1+\kappa_{g}\left(  f^{2}+s^{2}\right)  /4}
\tag{126}%
\end{equation}
where we define the geometrically induced fine structure function
$\alpha\equiv\kappa_{g}\left(  f^{2}+s^{2}\right)  /4$

\subsection{Primordial symmetries of standard model and torsion field}

In \cite{acs} the cross section for neutrino helicity spin flip obtained from
this type of $f(R;T)$ model of gravitation with dynamic torsion field
introduced by us in \cite{di4}, was phenomenologically analyzed using the
relation with the axion decay constant $f_{a}$ (Peccei-Quinn parameter) due
the energy dependence of the cross section, Consequently, the link with the
phenomenological energy/mass window was found from the astrophysical and high
energy viewpoints. The important point is that, in relation with the torsion
vector interaction Lagrangian, the $f_{a}$ parameter gives an estimate of the
torsion field strength that can variate with time within cosmological
scenarios as the described above, potentially capable of modifying the overall
leptogenesis picture, the magnetogenesis, the bariogenesis and also to obtain
some indication about the primordial (super) symmetry of the early universe.

In FRW\ scenario given here we saw that the torsion through its dual vector, namely:%

\begin{equation}
h^{0}=\frac{2}{5}\frac{\delta_{a}^{0}C_{\tau}}{a^{2}}d\tau\wedge e^{a}\text{ }
\tag{127}%
\end{equation}
goes as $\sim a^{-2}$ with $C_{\tau}$ a covariantly constant vector
field$\left(  \text{e.g.:}\overset{\circ}{\nabla}C_{\tau}=0\right)  $ that we
take of the form $C_{\tau}\sim\left(  \overset{\cdot}{\theta}+q_{\tau}\right)
$ (due the Hodge-de Rham decomposition of $h_{\mu},$ expression$(42))$where
$\theta$is a pseudoscalar field playing the obvious role of axion and
$q_{\tau}:$vector field linking $h^{0}$ with the magnetic field via the
equation of motion for the torsion$.$ Consequently, the torsion dual vector
$h^{i}$has the maximum value when the radius of the universe is $a_{\min}$ ,
e.g. $a_{\min}=a\left(  \tau_{0}\right)  $ increases to the maximum value the
spin-flip neutrino cross section and, for instance, the quantity of right
neutrinos compensating consequently the actual (e.g. $a_{today}=a\left(
\tau\right)  )$assymetry of the electroweak sector of the SM (see the
behaviour of $a$ in Fig.1). This fact indicates that the original symmetry
group contains naturally $SU_{R}\left(  2\right)  \times$ $SU_{L}\left(
2\right)  \times U\left(  1\right)  $ tipically inside GUT's structurally
based generally in $SO(10)$, $SU(5)$ or some exceptional groups as
$E(6)$,$E\left(  7\right)  ,etc$.%

\begin{figure}
[ptb]
\begin{center}
\includegraphics[
natheight=4.077600in,
natwidth=3.678100in,
height=4.1277in,
width=3.7256in
]%
{OQVLVJ00.wmf}%
\end{center}
\end{figure}

\bigskip

Also it is interesting to note that from the FRW\ line element written in
terms of the cosmic time the Hubble flow electromagnetic fields $E_{\mu}%
\equiv\left(  0,E_{i}\right)  =a^{-2}\left(  0,\partial_{\tau}A_{i}\right)  $
and $B_{\mu}\equiv\left(  0,B_{i}\right)  =a^{-2}\left(  0,\varepsilon
_{ijk}\partial_{j}A_{k}\right)  $%
\begin{align}
\overline{\nabla}\cdot\overline{E}+\left(  \frac{\alpha}{f}\overline{\nabla
}\theta+\overline{\Pi}\right)  \cdot\left(  a^{2}\overline{B}\right)   &
=0\tag{111}\\
\partial_{\tau}\left(  a^{2}\overline{E}\right)  -\overline{\nabla}%
\times\left(  a^{2}\overline{B}\right)   &  =\left(  \frac{\alpha}{f}%
\partial_{\tau}\theta+\Pi_{0}\right)  \left(  a^{2}\overline{B}\right)
-\left(  \frac{\alpha}{f}\overline{\nabla}\theta+\overline{\Pi}\right)
\times\overline{E}\tag{112}\\
\overline{\nabla}\cdot\overline{B}  &  =0\tag{113}\\
\partial_{t}\overline{B}  &  =-\overline{\nabla}\times\overline{E} \tag{114}%
\end{align}
where $\Pi_{\mu}\equiv f_{\mu}\left(  u_{\mu},\gamma^{5}b_{\mu},eA_{\mu
}.....\right)  $ is a vector function of physical entities as potential
vector, vorticity, angular velocity, axial vector etc etc. as described by
expression (42). In principle we can suppose that it is zero (low back
reaction \cite{kolb}) then
\begin{equation}
\overline{h}=\frac{\alpha}{f}\overline{\nabla}\theta,\text{
\ \ \ \ \ \ \ \ \ \ \ \ \ }h^{0}=\text{\ }\frac{\alpha}{f}\partial_{\tau
}\theta\text{\ } \tag{115}%
\end{equation}

being $\left[  \partial_{\tau}^{2}-\overline{\nabla}^{2}-\text{\ }\frac
{\alpha}{f}\partial_{\tau}\theta\overline{\nabla}\times\right]  \left(
a^{2}\overline{B}\right)  =0$ the second order equation for the magnetic field
that shows the chiral character of the plasma particles.

\subsection{Magnetogenesis and cosmic helicity}

Now we pass to see which role plays the torsion field in the magnetic field
generation in a FRW cosmology. Taking as the starting point the (hyper)
electrodynamic equations \cite{sha}and introducing a Fourier mode
decomposition $\overline{B}\left(  \overline{x}\right)  =\int d^{3}%
\overline{k}\overline{B}\left(  \overline{k}\right)  e^{-i\overline{k}%
\cdot\overline{x}}$ with $\overline{B}\left(  \overline{k}\right)
=h_{i}\overrightarrow{e}_{i}$ where $i=1,2$, $\overrightarrow{e}_{i}^{2}=1$,
$\overrightarrow{e}_{i}\cdot\overrightarrow{k}=\overrightarrow{e}_{1}%
\cdot\overrightarrow{e}_{2}=0$ the torsion-modified dynamical equations for
the expanding FRW become%

\begin{equation}
\overset{\cdot}{\overline{z}}+\left[  \left(  2\overset{\cdot}{a}+\frac{k^{2}%
}{\sigma}\right)  +\frac{ah^{0}\left\vert k\right\vert }{\sigma}\right]
\overline{z}=0 \tag{116}%
\end{equation}

\begin{equation}
\overset{\cdot}{z}+\left[  \left(  2\overset{\cdot}{a}+\frac{k^{2}}{\sigma
}\right)  -\frac{ah^{0}\left\vert k\right\vert }{\sigma}\right]  z=0 \tag{117}%
\end{equation}
where the magnetic field is written in terms of complex variable $z\left(
\overline{z}\right)  $ as%
\begin{align}
z  &  =h_{1}+ih_{2}\tag{118}\\
\overline{z}  &  =h_{1}-ih_{2} \tag{119}%
\end{align}
from equation (117) we see that the solution for $z$ namely:
\begin{equation}
z=z_{0}e^{-\left(  2a+\frac{k^{2}}{\sigma}\tau\right)  +\int\frac
{ah^{0}\left\vert k\right\vert }{\sigma}d\tau} \tag{120}%
\end{equation}
contains the instable mode in the sense of \cite{sha}$\frac{k}{\sigma}%
\tau<\int\frac{ah^{0}}{\sigma}d\tau.$ Consequently a defined polarization of
the magnetic field appear and from the dynamical equation for the torsion
field: $\nabla_{\left[  \mu\right.  }h_{\left.  \nu\right]  }=-\lambda
\widetilde{F}_{\mu\nu}$ that in this case we have
\begin{equation}
\nabla_{\left[  i\right.  }h_{\left.  \tau\right]  }=\nabla_{i}\left(
a^{-2}q_{\tau}\right)  =-\lambda B_{i} \tag{121}%
\end{equation}
that implies a relation between the vector part of the $h^{0}$ (namely
$q_{\tau})$ with the vector potential $A^{k}$ of the magnetic field as
follows:
\begin{equation}
\nabla_{i}q_{\tau}\approx-\lambda\varepsilon_{ijk}\nabla^{j}A^{k} \tag{122}%
\end{equation}
Consequently, the primordial magnetic field (or seed) would be connected in a
self-consistent way with the torsion field by means of the dual vector
$h_{0}.$It $(h_{\mu})$ in turn, would be connected phenomenologically with the
physical fields (matter) of theory through Hodge-de Rham decomposition
expression (42). We note from expression (120) that the pseudo-scalar (axion)
controls the stability, growth and dynamo effect but not the generation of the
magnetic field (primordial or seed) as is clear from expression (122) where
the (pseudo) -vector part of $h_{0}$ contributes directly to the generation of
the magnetic field as clearly given by eq. (121)

\subsection{Magnetogenesis and cosmic helicity II}

In the case to include the \textit{complete alpha term} given by equations
(92)\ and in the same analytical conditions (e.g.: Fourier decomposition) from
the previous paragraph, the torsion-modified dynamic equations for the
expanding FRW become%

\begin{equation}
\overset{\cdot}{\overline{z}}+\left[  \left(  2\overset{\cdot}{a}+\frac{k^{2}%
}{\sigma}\right)  +\frac{a\left\vert k\right\vert }{\sigma}\left(  h^{0}%
-\frac{\cos\alpha\left\vert \overline{j}_{gen}\right\vert }{\cos
\beta\left\vert \overline{z}\right\vert }\right)  \right]  \overline{z}=0
\tag{123}%
\end{equation}

\begin{equation}
\overset{\cdot}{z}+\left[  \left(  2\overset{\cdot}{a}+\frac{k^{2}}{\sigma
}\right)  -\frac{a\left\vert k\right\vert }{\sigma}\left(  h^{0}-\frac
{\cos\alpha\left\vert \overline{j}_{gen}\right\vert }{\cos\beta\left\vert
z\right\vert }\right)  \right]  z=0 \tag{124}%
\end{equation}
where in this case the magnetic field is written (by convenience) in terms of
complex variable $z\left(  \overline{z}\right)  $ as%
\begin{align}
z  &  =\left\vert z\right\vert e^{i\rho}\rightarrow\overset{\cdot}{z}=\left(
\overset{\cdot}{\left\vert z\right\vert }+i\overset{\cdot}{\rho}\left\vert
z\right\vert \right)  e^{i\rho}\tag{125}\\
\overline{z}  &  =\left\vert \overline{z}\right\vert e^{-i\rho}\rightarrow
\overset{\cdot}{\overline{z}}=\left(  \overset{\cdot}{\left\vert \overline
{z}\right\vert }-i\overset{\cdot}{\rho}\left\vert \overline{z}\right\vert
\right)  e^{-i\rho} \tag{126}%
\end{align}
From equation (124) we see that the solution for $z$ namely:
\begin{align}
z  &  =z_{0}\exp\left[  -\left(  2a+\frac{k^{2}}{\sigma}\tau\right)
+\int\frac{a\left\vert k\right\vert }{\sigma}\left(  h^{0}-\frac{\cos
\alpha\left\vert \overline{j}_{gen}\right\vert }{\cos\beta\left\vert
z_{0}\right\vert }\right)  d\tau\right] \tag{127}\\
\text{ with }z_{0}  &  =\left\vert z_{0}\right\vert e^{i\rho_{0}}\text{
\ \ \ \ \ }(\left\vert z_{0}\right\vert =const)\nonumber
\end{align}
contains the instable mode in the sense of \cite{sha} for example
$(117)\frac{k}{\sigma}\tau<\int\frac{a}{\sigma}\left(  h^{0}-\frac{\cos
\alpha\left\vert \overline{j}_{gen}\right\vert }{\cos\beta\left\vert
z\right\vert }\right)  d\tau.$ But now \textit{there are not a definite
polarization for the magnetic field}, but now all depends on the difference:%
\[
\int\frac{a}{\sigma}\left(  h^{0}-\frac{\cos\alpha\left\vert \overline
{j}_{gen}\right\vert }{\cos\beta\left\vert z_{0}\right\vert }\right)  d\tau
\]
Replacing explicitly $h_{\alpha}$ from the decomposition (42) we can see in a
clear form, the interplay between the physical entities, as the vortical and
magnetic helicities for example:
\[
\left(  \nabla_{0}\Omega+\frac{4\pi}{3}\left[  h_{M}+q_{s}n_{s}\overline
{u}_{s}\cdot\overline{B}\right]  +\gamma_{1}h_{V}+\gamma_{2}P_{0}\right)
-\frac{\cos\alpha\left\vert \overline{j}_{gen}\right\vert }{\cos
\beta\left\vert z_{0}\right\vert }%
\]
Now considering in $\left\vert \overline{j}_{gen}\right\vert $the fermionic
current $\underset{f}{\sum}q_{f}\overline{\psi}_{f}\gamma_{\mu}\psi_{f}$ ,
$\Omega$ as the axion $a$ , $\left\vert z_{0}\right\vert =\frac{\cos\beta
}{\cos\alpha}$and putting $\gamma_{2}=0$ we have an interesting expression:%
\[
\nabla_{0}a+\frac{4\pi}{3}\left[  h_{M}+q_{s}n_{s}\overline{u}_{s}%
\cdot\overline{B}\right]  +\gamma_{1}h_{V}=\left\vert \underset{f}{\sum}%
q_{f}\overline{\psi}_{f}\gamma_{\mu}\psi_{f}\right\vert
\]
The above expression it is very important because establishes the desired
connection between helicities, magnetic field and fermionic fields and axion.
We can order it as%
\[
\nabla_{0}a-\left\vert \underset{f}{\sum}q_{f}\overline{\psi}_{f}\gamma_{\mu
}\psi_{f}\right\vert =-\left[  \frac{4\pi}{3}\left(  h_{M}+q_{s}n_{s}%
\overline{u}_{s}\cdot\overline{B}\right)  +\gamma_{1}h_{V}\right]
\]
We now clearly see the link between the axion and the fermionic fields (the
dynamics of the interacting fields and the involved currents) in the LHS and
the macroscopic physical observables in the RHS giving an indication of the
origin of leptogenesis and bariogenesis in the context of this non Riemannian
gravitational model.

\subsection{Dark matter, energy condition and UFT model}

As is well know, in a wormhole solution energy conditions are always violated
in the standard general relativity. In the context of general relativity, this
fact is closely related to the necessity to introduce exotic matter through
the energy momentum tensor. Physically speaking, the observations of Type Ia
Supernova (SNIa), together with the cosmic microwave background radiation
(CMB)[ and the larger scale structure, suggest that the present universe is in
accelerating expansion, which needs something as dark energy with a negative
equation of state (e.g. phantom field, non-canonical dynamical terms, etc).
The simplest standard model introduces the cosmological constant term
$\Lambda$, which has a constant effective equation of state $w=-1$, and drive
the acceleration of the universe assuming the effective energy of the
$\Lambda$ term occupies$\sim$ $73\%$of the total energy (assuming also $\sim$
$23\%$ dark matter, $\sim4\%$ baryon matter and $\sim$ $10-5\%$ radiation)
constituting the $\Lambda$CDM model. This simple model satisfies more or less
all the cosmological observations but is still a phenomenological one. Also
the model suffers 2 important drawbacks: the`fine-tunning'\cite{ft} and the
`coincidence'\cite{coin} problems. In consequence other candidates as the dark
energy (especially the dynamical models) are required. Beyond the general
relativity, there are proposals in the literature where wormoholes of
different kinds \cite{bron} and other solutions with torsion were performed.
For example, the Einstein-Cartan model ECM is the simplest version of the
Poincare gauge theory of gravity (PGTG), in which the torsion is not dynamic
because the gravitational action is proportional to the curvature scalar of
Riemann-Cartan space-time (the ECT is a degenerate gauge theory in this
particular aspect): wormholes were treated for example in\cite{bron}.

The proposed model presented in this paper is purely geometric: no
energy-momentum tensor (EMT) is introduced. As have been seen,\ an effective
TEM (e.g.: eqs.(39-41)) is obtained when, from the general gravitational
equations, the standard Einstein tensor is isolated. In our case, the
effective TEM\ of the wormhole solution is diagonal (in the corresponding
coordinates) with the isotropic typical structure $\approx\mathbb{R}\otimes
SU\left(  2\right)  .$Consequently, we can proceed analyzing this effective
TEM\ (geometrical) by mean the standard energy conditions expressions
\cite{ellis}, namely:

i) \textit{WEC} \textit{(weak energy condition)}%
\[
\left.  T_{\mu\nu}\right\vert _{eff}\zeta^{\mu}\zeta^{\nu}\geqq0,\text{
\ \ \ \ \ \ \ \ \ \ \ }\left(  \zeta^{\nu}:\text{ any timelike vector}\right)
\Rightarrow\rho\geqq0,\rho+p_{k}\geqq0,\left(  k=1,2,3\right)
\]
guarantees that the energy density as measured by any local observer is nonnegative.

ii) \textit{DEC} \textit{(dominant energy condition)}%
\[
\left.  T_{00}\right\vert _{eff}\geqq\left\vert T_{ik}\right\vert
_{eff}\text{\ \ \ \ \ \ \ \ \ \ \ }\left(  \zeta^{\nu}:\text{ any timelike
vector}\right)  \Rightarrow\rho\geqq0,\rho+p_{k}\geqq0,\left(
i,k=1,2,3\right)
\]
includes \textit{WEC} and requires each each principal pressure never exceeds
the energy density which guarantees that the speed of sound cannot exceed the
light velocity c

iii) \textit{SEC} \textit{(strong energy condition)}%
\[
requires\rightarrow\rho+\sum p_{k}\geqq0
\]
and defines the sign of the acceleration due to gravity. In our case, the
wormhole solution presented in Fig 1), \textit{the condition iii) is fulfilled
jointly with conditions i) and ii)}. As we have made mention above, in
ref.\cite{bron} wormhole solutions with nondynamical torsion were constructed
in the context of the standard Einstein-Cartan model (ECM) fulfilling the
energy conditions also. The fundamental differences between the model in
ref.\cite{bron} and here are:

$\circ$In the case of \cite{bron} the energy conditions are fulfilled only for
particular values (local conditions or windows) of the parameters in the
introduced equation.

$\circ$In our case there are not free parameters but geometrically induced
functions mutually related.\textbf{ }Consequently, there are no parameters
that can be freely chosen but geometrically induced and mutually related
functions, so that the freedom to choose them independently is restricted:
e.g. once one of them is fixed, the others are automatically related to each
other by means of expressions of a dinamic character (like the analogue of
field b) or by means of the constraint given by $\Lambda$, etc (see Sections V
and VI). This important fact, which will be treated in a particular way in
another work\cite{dien}, would give an indication that the solutions could
have an overall character in our model.

\section{Discussion and perspectives}

In this paper we have introduced a simple geometric Lagrangian in the context
of a unified theory based on affine geometry. From the functional action
proposed, that is as square root or measure, the dynamic equations were
derived: an equation analogous to trace free Einstein equations $TFE$ and a
dynamic equation for the torsion (which was taken totally antisymmetric).
Although the aim of this paper was to introduce and to analize the model from
the viewpoint of previous research, we bring some new results and possible
explanations about the generation of primordial magnetic fields and the link
with the leptogenesis and baryogenesis. The physically admisible analysis of
the torsion vector $h_{\mu}$, from the point of view of the symmetries, has
allowed us to see how matter fields can be introduced in the model. These
fields include some dark matter candidates such as axion, right neutrinos and
Majorana. Also the vorticity can be included in the same way and, as the
torsion vector is connected to the magnetic field, both vorticity and magnetic
field can be treated with equal footing. The other point is that from the
wormhole soluton in a cosmological spacetime with torsion we show that
primordial cosmic magnetic fields can be originated by the dual torsion field
$h_{\mu}$ being the axion field contained in $h_{\mu}$, the responsible to
control the dynamics and stability of the cosmic magnetic field, but is not
responsible of the magnetogenesis itself. Also the energy conditions in the
wormhole solution are fulfilled. The last important point to highlight is that
the dynamic \ \ torsion field $h_{\mu}\,\ $acts as mechanism of the reduction
of an original (early, primordial) GUT\ (Grand Unified Theory) symmetry of the
universe containing $\sim$SU$\left(  3\right)  \times$SU$\left(  2\right)
_{R}\times$SU$\left(  2\right)  _{L}\times$U$\left(  1\right)  $
to\ SU$\left(  3\right)  \times$SU$\left(  2\right)  _{L}\times$U$\left(
1\right)  $ today. Consequently, the GUT candidates are $SO(10)$, $SU(5)$ or
some exceptional groups as $E(6)$,$E\left(  7\right)  $ for example.

\section{Acknowlegments}

I am very grateful to the JINR Directorate and the BLTP for his hospitality
and CONICET-ARGENTINA for financial support. Many thanks are given to Max
Camenzind for his comments about gauge theories and gravitation, Brett Mc
Innes that bring me important references and to Andrej Arbuzov for his useful
insigts and comments. This work is devoted to the memory of Professor Victor
Nicolaievich Pervushin, active and enthusiastic scientist that suddenly pass
away this year .

\section{Appendix I}

We must to remind that the model where the interaction arises is based on a
pure affine geometrical construction wher the geometrical Lagrangian of the
theory contains dynamically the generalized curvature $\mathcal{R=}%
\det(\mathcal{R}_{\ \mu}^{a})$, namely
\[
L_{g}=\sqrt{det\mathcal{R}_{\ \mu}^{a}\mathcal{R}_{a\nu}}=\sqrt{detG_{\mu\nu}}%
\]
characterizing a higher dimensional group manifold e.g: SU(2,2). Then, after
the breaking of the symmetry, typically from the conformal to the Lorentz
group e.g: $SU(2,2)\rightarrow SO\left(  1,3\right)  ,$ the generalized
curvature becomes to
\[
\mathcal{R}_{\ \mu}^{a}=\lambda\left(  e_{\ \mu}^{a}+f_{\ \mu}^{a}\right)
+R_{\ \mu}^{a}\qquad\left(  M_{\mu}^{a}\equiv e^{a\nu}M_{\nu\mu}\right)
\]
taking the original Lagrangian $L_{g}$ the following form: $L_{g}%
\rightarrow\sqrt{Det\left[  \lambda^{2}\left(  g_{\mu\nu}+f_{\ \mu}^{a}%
f_{a\nu}\right)  +2\lambda R_{\left(  \mu\nu\right)  }+2\lambda f_{\ \mu}%
^{a}R_{[a\nu]}+R_{\ \mu}^{a}R_{a\nu}\right]  },$reminiscent of a nonlinear
sigma model or M-brane. Notice that $f_{\ \mu}^{a}$ , in a sharp contrast with
the tetrad field $e_{\ \mu}^{a}$, carries the symmetry $e_{a\mu}f_{\ \nu}%
^{a}=f_{\mu\nu}=-f_{\nu\mu}.$-- see [38,39] \cite{di}\cite{di1}\cite{di2}%
\cite{di3}for more mathematical and geometrical details of the theory.

Consequently the generalized Ricci tensor splits into a symmetric and
antisymmetric part, namely:
\[
R_{\mu\nu}=\overset{R_{\left(  \mu\nu\right)  }}{\overbrace{\overset{\circ}%
{R}_{\mu\nu}-T_{\mu\rho}^{\ \ \ \alpha}T_{\alpha\nu}^{\ \ \ \rho}}}%
+\overset{R_{\left[  \mu\nu\right]  }}{\overbrace{\overset{\circ}{\nabla
}_{\alpha}T_{\mu\nu}^{\ \ \ \alpha}}}%
\]
where $\overset{\circ}{R}_{\mu\nu}$ is the general relativistic Ricci tensor
constructed with the Christoffel connection, $T_{\mu\rho}^{\ \ \ \alpha
}T_{\alpha\nu}^{\ \ \ \rho}$ is the quadratic term in the torsion field and
the antisymmetric last part $\overset{\circ}{\nabla}_{\alpha}T_{\mu\nu
}^{\ \ \ \alpha}$ is the divergence of the totally antisymmetric torsion field
that introduce its dynamics in the theory. From a theoretical point of view
our theory containing a dynamical totally antisymmetric torsion field is
comparable to that of Kalb-Ramond in string or superstring theory
\textbf{\cite{gsw}} but in our case all: energy, matter and interactions are
geometrically induced.

Notice that $^{\ast}f_{\mu\nu}$in $L_{g}$ must be proportional to the physical
electromagnetic field, namely $jF_{\mu\nu}$ where the parameter $j$
homogenizes the units such that the combination $g_{\mu\nu}+jF_{\mu\nu}$ has
the correct sense. We will not go into details but the great advantage of the
model is that it is purely geometric without energy-momentum tensor added by hand.

\section{Appendix II}

On the g-variation:

from%

\begin{equation}
\mathcal{L}=\sqrt{\left\vert g\right\vert }\sqrt{\det\left(  \alpha
\lambda\right)  }\sqrt{1+\frac{1}{2b^{2}}F_{\mu\nu}F^{\mu\nu}-\frac{1}%
{16b^{4}}\left(  F_{\mu\nu}\widetilde{F}^{\mu\nu}\right)  ^{2}}\equiv
\sqrt{\left\vert g\right\vert }\sqrt{\det\left(  \alpha\lambda\right)
}\mathbb{R} \tag{1}%
\end{equation}
where%
\begin{gather}
b=\frac{\alpha}{\beta}=\frac{1+\left(  R_{S}/4\lambda\right)  }{1+\left(
R_{A}/4\lambda\right)  },\tag{2}\\
R_{S}=g^{\alpha\beta}R_{\alpha\beta},\tag{3}\\
R_{A}=f^{\alpha\beta}R_{\alpha\beta},\text{ \ \ } \tag{4}%
\end{gather}
and $\lambda$ arbitrary constant. Knowing that, in the metrical case we have
as usual procedure:%
\begin{equation}
\delta_{g}\mathcal{L}=\left[  \delta\left(  \sqrt{\left\vert g\right\vert
}\sqrt{\det\left(  \alpha\lambda\right)  }\right)  \mathbb{R}+\sqrt{\left\vert
g\right\vert }\sqrt{\det\left(  \alpha\lambda\right)  }\delta\mathbb{R}%
\right]  \tag{5}%
\end{equation}%
\begin{equation}
\delta\left(  F_{\mu\nu}F^{\mu\nu}\right)  =2F_{\mu\lambda}F_{\nu}^{\ \lambda
}\delta g^{\mu\nu} \tag{6}%
\end{equation}

\begin{equation}
\delta\left(  \widetilde{F}_{\mu\nu}F^{\mu\nu}\right)  =\left(  -\frac{1}%
{2}\widetilde{F}_{\eta\rho}F^{\eta\rho}g_{\mu\nu}+4\widetilde{F}_{\mu\rho
}F_{\nu}^{\ \rho}\right)  \delta g^{\mu\nu} \tag{7}%
\end{equation}
then%
\begin{gather}
\left[  2R_{\left(  \alpha\beta\right)  }-\frac{g_{\alpha\beta}}{2}%
R_{s}\right]  \mathbb{R}=\frac{R_{s}}{\mathbb{R}\alpha^{2}}\left[
F_{\alpha\lambda}F_{\beta}^{\ \lambda}+\frac{1}{2b^{2}}F_{\mu\nu}\widetilde
{F}^{\mu\nu}\left(  \frac{F_{\eta\rho}\widetilde{F}^{\eta\rho}}{8}%
g_{\alpha\beta}-F_{\alpha\lambda}\widetilde{F}_{\beta}^{\ \lambda}\right)
\right]  -F_{\mu\nu}F^{\mu\nu}R_{\left(  \alpha\beta\right)  }+\tag{8}\\
+\frac{R_{\left(  \alpha\beta\right)  }}{\mathbb{R}\alpha^{2}}\left[
\frac{\left(  F_{\eta\rho}\widetilde{F}^{\eta\rho}\right)  ^{2}}%
{4\mathbb{R}^{2}b^{2}}-F_{\mu\nu}F^{\mu\nu}\right]  +4\lambda\left[
g_{\alpha\beta}+\frac{1}{\mathbb{R}^{2}\alpha^{2}}\left(  F_{\alpha\lambda
}F_{\beta}^{\ \lambda}+\frac{F_{\mu\nu}\widetilde{F}^{\mu\nu}}{2b^{2}}\left(
\frac{F_{\eta\rho}\widetilde{F}^{\eta\rho}}{8}g_{\alpha\beta}-F_{\alpha
\lambda}\widetilde{F}_{\beta}^{\ \lambda}\right)  \right)  \right]  ,\nonumber
\end{gather}

\bigskip%
\begin{align}
R_{\left(  \alpha\beta\right)  }-\frac{g_{\alpha\beta}}{4}R_{s}  &
=\frac{R_{s}}{2\mathbb{R}^{2}\alpha^{2}}\left[  F_{\alpha\lambda}F_{\beta
}^{\ \lambda}-F_{\mu\nu}F^{\mu\nu}\frac{R_{\left(  \alpha\beta\right)  }%
}{R_{s}}\right]  +\tag{9}\\
&  +\frac{R_{s}}{4\mathbb{R}^{2}\alpha^{2}b^{2}}\left[  F_{\mu\nu}%
\widetilde{F}^{\mu\nu}\left(  \frac{F_{\eta\rho}\widetilde{F}^{\eta\rho}}%
{8}g_{\alpha\beta}-F_{\alpha\lambda}\widetilde{F}_{\beta}^{\ \lambda}\right)
+\frac{F_{\eta\rho}\widetilde{F}^{\eta\rho}}{2}\frac{R_{\left(  \alpha
\beta\right)  }}{R_{s}}\right]  +\nonumber\\
&  +2\lambda\left[  g_{\alpha\beta}+\frac{1}{\mathbb{R}^{2}\alpha^{2}}\left(
F_{\alpha\lambda}F_{\beta}^{\ \lambda}+\frac{F_{\mu\nu}\widetilde{F}^{\mu\nu}%
}{2b^{2}}\left(  \frac{F_{\eta\rho}\widetilde{F}^{\eta\rho}}{8}g_{\alpha\beta
}-F_{\alpha\lambda}\widetilde{F}_{\beta}^{\ \lambda}\right)  \right)  \right]
,\nonumber
\end{align}

\section{Appendix III}

Some remarks on the general Hodge-de Rham decomposition of $h=h_{\alpha
}dx^{\alpha}$

\begin{theorem}
if $h=h_{\alpha}dx^{\alpha}$ $\notin F^{\prime}\left(  M\right)  $ is a 1-form
on $M$, then there exist a zero-form $\Omega$, a 2-form $\alpha=A_{\left[
\mu\nu\right]  }dx^{\mu}\wedge dx^{\nu}$ and an harmonic 1-form $q=q_{\alpha
}dx^{\alpha}$ on $M$ that%
\begin{equation}
h=d\Omega+\delta\alpha+q\rightarrow h_{\alpha}=\nabla_{\alpha}\Omega
+\varepsilon_{\alpha}^{\beta\gamma\delta}\nabla_{\beta}A_{\gamma\delta
}+q_{\alpha} \tag{5}%
\end{equation}

\end{theorem}

Notice that even if is not harmonic, and assuming that $q_{\alpha}$ is a polar
vector, an axial vector can be added such that the above expression takes the
form%
\begin{equation}
h_{\alpha}=\nabla_{\alpha}\Omega+\varepsilon_{\alpha}^{\beta\gamma\delta
}\nabla_{\beta}A_{\gamma\delta}+\varepsilon_{\alpha}^{\beta\gamma\delta
}M_{\beta\gamma\delta}+q_{\alpha} \tag{6}%
\end{equation}
where $M_{\beta\gamma\delta}$ is a completely antisymmetric tensor.(of such a
manner that $\varepsilon_{\alpha}^{\beta\gamma\delta}M_{\beta\gamma\delta}$
$\equiv\gamma^{5}b_{\alpha}$is an axial vector).

Consequently, we know that in unified theories where we are not able to deal
with energy-momentum tensor, the fields and they interactions are effectively
restricted due the same geometrical framework: the spacetime itself. This fact
permits us to rewrite (14) considering the physical quantities of interest:%

\[
h_{\alpha}=\nabla_{\alpha}\Omega+\varepsilon_{\alpha}^{\beta\gamma\delta
}\nabla_{\beta}A_{\gamma\delta}+\gamma^{5}b_{\alpha}+\left(  P_{\alpha
}-eA_{\alpha}\right)
\]

\section{Appendix IV}

\subsection{Electrodynamical equations in 3+1}

the starting point will be the line element in $3+1$ splitting\cite{kt}%
\cite{kt1}: the 4-dimensional spacetime is split into 3-dimensional space and
1-dimensional time to form a foliation of 3-dimensional spacelike
hypersurfaces. The metric of the spacetime is consequently, given by%
\[
ds^{2}=-\alpha^{2}dt^{2}+\gamma_{ij}\left(  dx^{i}+\beta^{i}dt\right)  \left(
dx^{j}+\beta^{j}dt\right)
\]
where $\gamma_{ij}$ is the metric of the 3-dimensional hypersurface,$\alpha$is
the lapse function, and $\beta^{i}$ is the shift function. At every spacetime
point, a fiducial observer (FIDO) is introduced in such a way that his
corresponding world-line is perpendicular to the hypersurface where he is stationary.

His FIDO 4-vector velocity is then given by%
\[
U^{\mu}=\frac{1}{\alpha}\left(  1,-\beta^{i}\right)  ,U_{\mu}=\left(
-\alpha,0,0,0\right)
\]
one deals with the physical quantities defined on the 3-dimensional
hypersurface as measured by the FIDO. For example, the electric field and the
magnetic field are defined with the help of the $U^{\mu}$ respectively, by%
\begin{align*}
E^{\mu}  &  =F^{\mu\nu}U_{\nu}\\
B^{\mu}  &  =-\frac{1}{2\sqrt{-g}}\varepsilon^{\mu\nu\rho\sigma}U_{\nu}%
F_{\mu\nu}%
\end{align*}
notice that the zero components are null: $E^{0}=B^{0}=0$. Also, the 4-current
$J^{\mu}$ can be similarly decomposed as%
\[
J^{\mu}=\rho_{e}U^{\mu}+j^{\mu}%
\]
where we defined%
\begin{align*}
\rho_{e}  &  =-J^{\mu}U_{\mu}\\
j^{\mu}  &  =J^{\mu}+J^{\nu}U_{\nu}U^{\mu}%
\end{align*}

then $j^{0}=0$. So that $j$, $E$ and $B$ can be treated as 3-vectors in
spacelike hypersurfaces. In terms of these 3-vectors the Maxwell eqs. can be
written as
\begin{align*}
\nabla\cdot E  &  =4\pi\rho_{e}\\
\nabla\cdot B  &  =0\\
\nabla\times\left(  \alpha E\right)   &  =-(\partial_{t}-\mathcal{L}%
_{\mathcal{\beta}})B\\
&  =-\partial_{0}B+\left(  \beta\cdot\nabla\right)  B-\left(  B\cdot
\nabla\right)  \beta\\
\nabla\times\left(  \alpha B\right)   &  =-(\partial_{t}-\mathcal{L}%
_{\mathcal{\beta}})E+4\pi\alpha j\\
&  =\partial_{0}E-\left(  \beta\cdot\nabla\right)  E+\left(  E\cdot
\nabla\right)  \beta+4\pi\alpha j
\end{align*}
The derivatives in these equations are covariant derivatives with respect to
the metric of the absolute space $\gamma_{ij}$ being $\mathcal{L}%
_{\mathcal{\beta}}$ the Lie derivative operator geometrically defined as:
$\mathcal{L}_{\mathcal{\beta}}V=d\left(  i_{\beta}\cdot V\right)  $ with $V$ a
vector field.

ZAMOs observers
\[
U=\frac{1}{\alpha}\left(  \partial_{t}-\beta^{i}e_{i}\right)
\]
in the Boyer-Lindquist coordinates we have $e_{r},e_{\theta}$ and $e_{\varphi
}=\frac{1}{\sqrt{g_{\varphi\varphi}}}\partial_{\varphi}.$ The plasma
$4-$velocity (medium) $u$ can be expressed as $u=\gamma\left(  U+\overline
{v}\right)  $ where $\overline{v}$ is the plasma $3-$velocity with respect to
the ZAMOs.

\end{document}